\expandafter\def\csname ver@fixltx2e.sty\endcsname{}
\PassOptionsToPackage{pdfpagelabels=false}{hyperref}
\documentclass[fleqn,usenatbib,useAMS]{mnras}
\usepackage[T1]{fontenc}
\usepackage{graphicx}	
\usepackage{amsmath}	
\usepackage{amssymb}
\usepackage{newtxtext,newtxmath}
\usepackage{multicol}   
\usepackage{bm}		
\usepackage{pdflscape}
\usepackage{ae,aecompl}
\usepackage{natbib}

\title[Bar rotation rate as dark matter diagnostic]{The bar rotation rate as a diagnostic of dark matter content in the centre of disc galaxies}
\author[C. Buttitta et al.]{C. Buttitta,$^{1}$ 
E.~M. Corsini,$^{1,2}$
J.~A.~L. Aguerri,$^{3}$
L. Coccato,$^{4}$
L. Costantin,$^{5}$
V. Cuomo,$^{6}$
\newauthor{V.~P. Debattista,$^{7}$
L. Morelli,$^{6}$ 
and A. Pizzella$^{1,2}$}
\\
$^{1}$Dipartimento di Fisica e Astronomia ``G. Galilei'', Universit\`a di Padova, vicolo dell'Osservatorio 3, I-35122 Padova, Italy\\
$^{2}$INAF - Osservatorio Astronomico di Padova, vicolo dell'Osservatorio 2, I-35122 Padova, Italy \\
$^{3}$Instituto de Astrof\'\i sica de Canarias, Calle V\'\i a L\'actea s/n, E-38205 La Laguna, Tenerife, Spain\\
$^{4}$European Southern Observatory, Karl-Schwarzschild-Strasse 2, D-85748 Garching, Germany\\
$^{5}$Centro de Astrobiolog\'{\i}a (CSIC-INTA), Ctra de Ajalvir km 4, Torrej\'on de Ardoz, 28850, Madrid, Spain\\
$^{6}$Instituto de Astronom\'{\i}a y Ciencias Planetarias, Universidad de Atacama, Avenida Copayapu 485, Copiap\'o, Chile\\
$^{7}$Jeremiah Horrocks Institute, University of Central Lancashire, PR1 2HE Preston, UK\\
}

\date{}

\pubyear{2022}

\begin{document}
\label{firstpage}
\pagerange{\pageref{firstpage}--\pageref{lastpage}}
\maketitle

% Abstract of the paper
\begin{abstract}
We investigate the link between the bar rotation rate and dark matter content in barred galaxies by concentrating on the cases of the lenticular galaxies NGC\,4264 and NGC\,4277. These two gas-poor galaxies have similar morphologies, sizes, and luminosities. But, NGC\,4264 hosts a fast bar, which extends to nearly the corotation, while the bar embedded in NGC\,4277 is slow and falls short of corotation. 
We derive the fraction of dark matter $f_{\rm DM, bar}$ within the bar region from Jeans axisymmetric dynamical models by matching the stellar kinematics obtained with the MUSE integral-field spectrograph and using SDSS images to recover the stellar mass distribution. We build mass-follows-light models as well as mass models with a spherical halo of dark matter, which is not tied to the stars. 
We find that the inner regions of NGC\,4277 host a larger fraction of dark matter ($f_{\rm DM, bar}\,=\,0.53\pm0.02$) with respect to NGC\,4264 ($f_{\rm DM, bar}\,=\,0.33\pm0.04$) in agreement with the predictions of theoretical works and the findings of numerical simulations, which have found that fast bars live in baryon-dominated discs, whereas slow bars experienced a strong drag from the dynamical friction due to a dense DM halo. This is the first time that the bar rotation rate is coupled to $f_{\rm DM, bar}$ derived from dynamical modelling.
\end{abstract}

% Select between one and six entries from the list of approved keywords.
% Don't make up new ones.

%%%%%%%%%%%%%%%%%%%%%%%%%%%%%%%%%%%%%%%%%%%%

%%%%%%%%%%%%%% BODY OF PAPER %%%%%%%%%%%%%%%

\section{Introduction}
\label{sec:intro}

About two-thirds of disc galaxies, including the Milky Way, have a bar which is tumbling at the centre of the disc \citep{Aguerri2009, Buta2015}. The bar is an efficient agent for redistributing the stars by exchanging angular momentum, energy, and mass among the different galactic components including the dark matter (DM) halo \citep{Athanassoula2013, Sellwood2014}.

The main properties of bars are the radius $R_{\rm bar}$, which defines the elongation of the stellar orbits in the bar, strength $S_{\rm bar}$, which quantifies the non-axisymmetric contribution of the bar to the gravitational potential, pattern speed $\Omega_{\rm bar}$, which is the angular frequency of the bar figure rotation, and the rotation rate $\mathcal{R}$. The latter is defined as the dimensionless ratio between the length of the corotation radius $R_{\rm cor}$, where stars circle the galactic centre at $\Omega_{\rm bar}$, and bar radius. The rotation rate does not depend on galaxy distance and distinguishes between fast ($1\leq \mathcal{R}\leq1.4$) and slow bars ($\mathcal{R}>1.4$) \citep{Athanassoula1992, Debattista2000}.

The two main mechanisms which trigger the formation of a bar are internal gravitational instabilities of the stellar disc \citep{Sellwood1981} and external tidal interactions \citep{Noguchi1987}. Spontaneously-formed bars are usually thin, long, and fast \citep{Athanassoula2013}, whereas tidally-induced bars are thick, short, and slow \citep{Martinez-Valpuesta2017}. The bar properties evolve with time reshaping the morphology, orbital structure, mass distribution, star formation, and stellar population properties of their host galaxies \citep{Laurikainen2007, Fragkoudi2016}. Once formed, the bar becomes longer and stronger and it slows down on timescales, which depend on the DM content in the disc region \citep{Debattista1998, Debattista2000, Athanassoula2002, Petersen2019}.

\begin{figure*}
\centering
\includegraphics[scale=0.32]{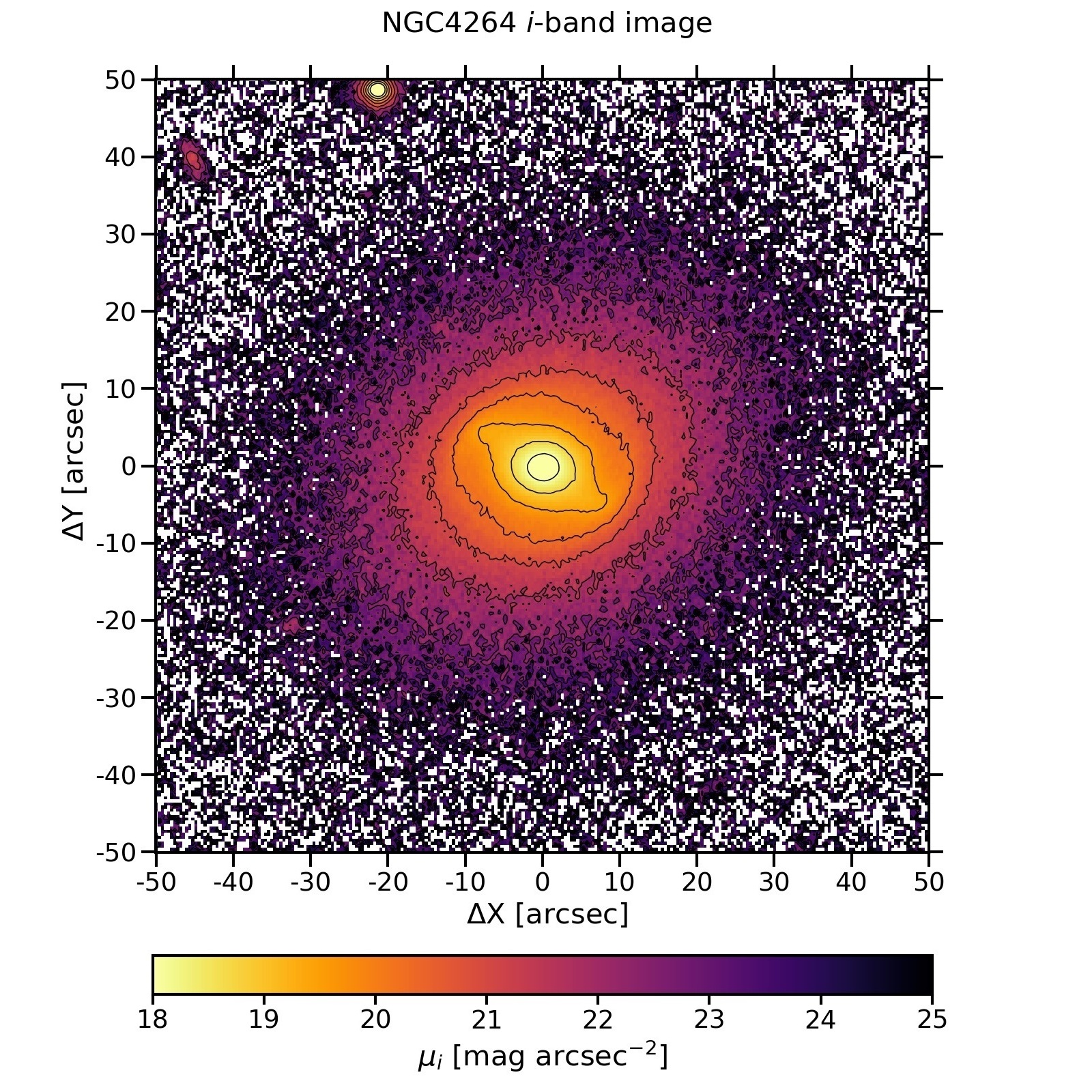}
\includegraphics[scale=0.32]{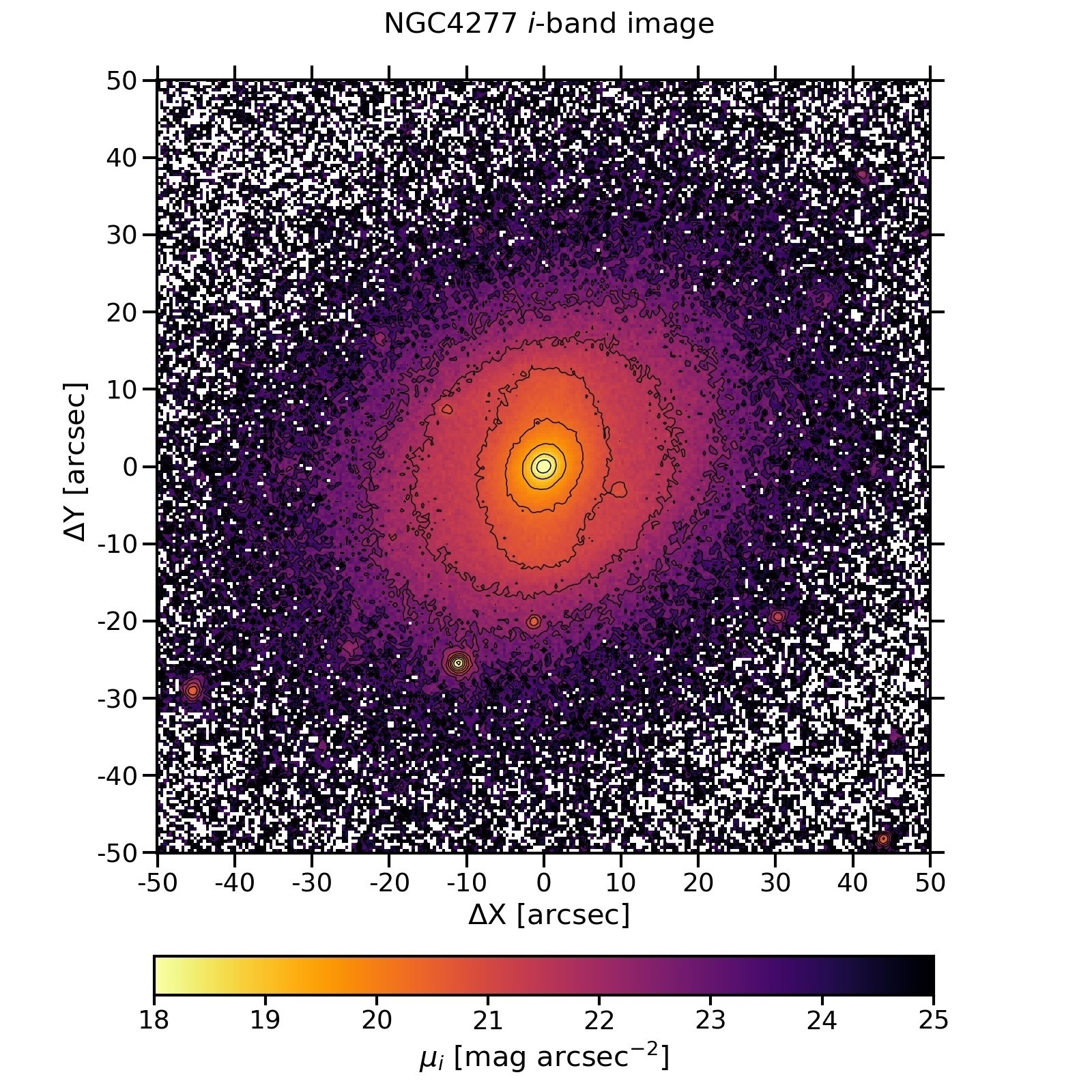}
\caption{SDSS $i$-band image of NGC\,4264 (left panel) and NGC\,4277 (right panel). Some reference isophotes, spaced by 0.5 mag\,arcsec$^{-2}$, are over-plotted with black lines. The FOV is $1.7 \times 1.7$ arcmin$^2$ and is oriented with North up and East left.}
\label{fig:SDSS_images}
\end{figure*}

In the last decade, the systematic investigation of the pattern speeds of large samples of barred galaxies with integral-field spectroscopy has shown that almost all bars are fast \citep{Aguerri2015, Guo2019, Cuomo2020, Garma-Oehmichen2020, Garma-Oehmichen2022} confirming the early findings based on long-slit spectroscopy of few selected objects \citep{Corsini2011}. This supports the idea that the central regions of lenticular and spiral galaxies host maximal (or nearly maximal) stellar discs with a low content of DM. 

These observationally-driven findings are in conflict with the predictions of some hydro-dynamical cosmological simulations, for which galaxies are embedded in centrally-concentrated DM halos required by the $\Lambda$CDM paradigm. \citet{Algorry2017} measured the bar properties in the galaxies extracted from the EAGLE simulation \citep{EAGLE2015} and found a reasonable agreement with the bar radii and strengths measured for real galaxies. However, the simulated bars experienced an intense slowdown due to the dynamical friction of the DM halo and many of them ended up slow at $z\sim0$. Similarly, \citet{Roshan2021} found that the bars in the TNG50 simulation \citep{Pillepich2018, Nelson2018} are much slower ($\mathcal{R}>1.9$) with respect to the observed ones. More recently, the discrepancy between observations and simulations has been attenuated by the findings of \citet{Fragkoudi2021} and \citet{Marioni2022}. \citet{Fragkoudi2021} analysed the barred galaxies in the AURIGA simulation suite \citep{AURIGA2017} and showed that they have fast bars because they are more baryon-dominated with respect to those in the TNG simulation. \citet{Marioni2022} investigated the evolution of barred galaxies in the CLUES simulation \citep{CLUES2010}, which have shorter, but not slower, bars with respect to their observed counterparts. A possible explanation for these findings could reside in the different ingredients of the simulations so far analysed, including the resolution of the simulation and the gas fraction, disc thickness, stellar and AGN feedback, and baryonic content of the simulated galaxies.

Dynamical models of barred galaxies with accurate measurements of $\Omega_{\rm bar}$ and $\mathcal{R}$ are needed to rigorously test the predictions of numerical simulations regarding the bar properties as a function of gas content, luminosity and DM distribution. In this paper, we derive the mass distribution of two barred galaxies, NGC\,4264 and NGC\,4277, for which the values of $\Omega_{\rm bar}$ are amongst the best-constrained ones ever obtained with direct measurements \citep{TW1984}. This will allow us to investigate the link between $\mathcal{R}$ and the DM content in the bar region because NGC\,4264 hosts a fast bar ($\mathcal{R}=0.9\pm0.2$, \citealt{Cuomo2019}) while the bar in NGC\,4277 is slow ($\mathcal{R}=1.8^{+0.5}_{-0.3}$, \citealt{Buttitta2022}). We aim at understanding whether a larger value of $\mathcal{R}$ results from the effective bar braking due to the dynamical friction exerted by the DM halo and therefore is a diagnostic of a large content of DM in the bar region. 

The paper is organised as follows. We present the main properties of the two galaxies and their bars in Secs.~\ref{sec:galaxy_properties} and \ref{sec:bar_properties}, respectively. We discuss the choice and application of the dynamical model in Sec.~\ref{sec:method}. We present our results and their implications in Secs.~\ref{sec:results} and \ref{sec:conclusion}, respectively. 

\section{Main properties of NGC~4264 and NGC~4277}
\label{sec:galaxy_properties}

NGC\,4264 and NGC\,4277 are two early-type disc galaxies classified as SB0$^+$(rs) and SAB(rs)0/a respectively  by \citet[][hereafter RC3]{RC3}. They have a quite similar morphology (Fig.~\ref{fig:SDSS_images}) with a well-defined bar surrounded by a pseudo-ring and oriented at an intermediate position angle with respect to the major and minor axes of the disc ($|{\rm PA_{bar}}-{\rm PA_{disc}}|\sim50^\circ$). The bar region appears to be mostly free of dust and star formation and the disc has an intermediate inclination ($i_{\rm disc}\sim40^\circ$). NGC\,4264 and NGC\,4277 have similar luminosity and size, as calculated from the apparent corrected magnitude $B_{\rm T}^0$ (RC3) and galaxy diameters $D_{25}$ and $d_{25}$ (RC3) by adopting the distance obtained from the radial velocity with respect to the cosmic microwave background reference frame \citep{Fixsen1996}. NGC\,4264 and NGC\,4277 are located behind the Virgo cluster. They likely form an interacting couple with the giant elliptical galaxy NGC\,4261 \citep{Schmitt2001} and with the spiral galaxy NGC\,4273 \citep{Kim2014}, respectively. The main properties of both galaxies are given in Table~\ref{tab:prop_galaxies}.

\begin{table}
\centering
\renewcommand{\tabcolsep}{0.15cm}
\renewcommand{\arraystretch}{1.05}
\begin{tabular}{l c l  c  c}
\hline
     & Property                 &                      & NGC\,4264           & NGC\,4277            \\
\hline
\hline
(1)  & Morph. Type              &                      & SB0$^+$(rs)         & SAB(rs)0/a           \\
(2)  & $M^0_{B_{\rm T}}$        & [mag]                & $-19.20$            & $-19.27$             \\
(3)  & $D$                      & [Mpc]                & $38.0\pm2.7$        & $33.8\pm2.4$         \\
(4)  & $D_{25}\,\times\,d_{25}$ & [kpc$^2$]            & $10.8\,\times\,8.8$ & $10.3\,\times\,8.6$  \\
(5)  & PA$_{\rm disc}$          & [$^{\circ}$]         & $114.0\pm1.2$       & $123.3\pm0.3$        \\
(6)  & $i_{\rm disc}$           & [$^{\circ}$]         & $36.7\pm0.7$        & $40.7\pm0.7$         \\
\hline
(7)  & PA$_{\rm bar}$           & [$^{\circ}$]         & $56.4\pm0.1$        & $175.59\pm0.04$      \\         
(8)  & $R_{\rm bar}$            & [kpc]                & $3.2\pm0.5$         & $3.2^{+0.9}_{-0.6}$  \\
(9)  & $S_{\rm bar}$            &                      & $0.31\pm0.04$       & $0.21\pm0.02$        \\
(10) & $V_{\rm circ}$           & [km s$^{-1}$]        & $189\pm10$          & $148\pm5$            \\
(11) & $\Omega_{\rm bar}$   & [km s$^{-1}$ kpc$^{-1}$] & $71\pm4$            & $25\pm3$             \\
(12) & $R_{\rm cor}$            & [kpc]                & $2.8\pm0.2$         & $6.0\pm0.9$          \\
(13) & $\mathcal{R}$            &                      & $0.9\pm0.2$         & $1.8^{+0.5}_{-0.3}$  \\
\hline 
\end{tabular}
\caption{Galaxy and bar properties of NGC\,4264 and NGC\,4277. (1): Morphological type from RC3. (2): Total absolute magnitude from $B_{\rm T}^0$ in RC3. (3): Distance calculated from the radial velocity with respect to the cosmic microwave background reference frame \citep{Fixsen1996} and assuming $H_0=75$ km s$^{-1}$ Mpc$^{-1}$, $\Omega_{\rm m} = 0.308$, and $\Omega_{\Lambda}=0.692$. (4): Major and minor diameters of the isophote with surface brightness $\mu_{B}=25$ mag arcsec$^{-2}$ from RC3. (5): Disc position angle from the isophotal analysis. (6): Disc inclination from the isophotal analysis assuming an infinitesimally thin disc. (7): Bar position angle from the photometric decomposition. (8): Bar radius. (9): Bar strength. (10): Circular velocity from the stellar streaming motion corrected for asymmetric drift. (11): Bar pattern speed. (12): Corotation radius. (13): Bar rotation rate.}
\label{tab:prop_galaxies}
\end{table}

\section{Properties of the bars}
\label{sec:bar_properties}

\citet{Cuomo2019} and \citet{Buttitta2022} analysed the surface photometry and integral-field spectroscopy of NGC\,4264 and NGC\,4277, respectively, to characterise the properties of their bars. They measured the bar radius and strength from the surface photometry obtained from broad-band imaging of the Sloan Digital Sky Survey (SDSS). They derived the bar pattern speed from the stellar kinematics obtained from integral-field spectroscopy performed with the Multi Unit Spectroscopic Explorer (MUSE) at the Very Large Telescope (VLT). They also estimated the co-rotation radius from the circular velocity, which they constrained by correcting the stellar streaming motions for asymmetric drift. Finally, they derived the bar rotation rate. Here, we provide a concise description of the acquisition and analysis of the photometric and kinematic data and a summary of the results. The properties of the bars of both galaxies are given in Table~\ref{tab:prop_galaxies}.

\subsection{Bar radius and strength}
\label{sec:length_strength}

\citet{Cuomo2019} and \citet{Buttitta2022} analysed the flux-calibrated $i$-band images of both galaxies available in the science archive of the SDSS Data Release 14 \citep{SDSSDR14}. They were obtained with a spatial sampling of $0.3961$ arcsec pixel$^{-1}$, total exposure time of 53.9 s, and seeing ${\rm FWHM}\sim1.5\,{\rm arcsec}$. The images were sky subtracted and trimmed selecting a field of view (FOV) of $800\,\times\,800$ pixel$^2$ centred on the galaxies corresponding to $5.3\,\times\,5.3$ arcmin$^2$. 

The isophotal analysis recovered the geometric parameters PA$_{\rm disc}$ and $i_{\rm disc}$ of the galaxy disc, which were adopted to deproject the galaxy image. The photometric decomposition was performed to estimate the position angle, PA$_{\rm bar}$, of the bar and its contribution to the galaxy surface brightness. 

For both galaxies, the radius $R_{\rm bar}$ of the bar was derived as the mean of the estimates obtained from the analysis of the radial profile of the position angle of the interpolated isophotes on the deprojected image as in \citet{Aguerri2003}, of the intensity contrast between the bar and interbar regions following \citet{Aguerri2000}, and of the photometric decomposition adopting a Ferrers bar as in \citet{MendezAbreu2017}. The strength $S_{\rm bar}$ of the bar was derived as the mean of the values obtained from the Fourier analysis of the deprojected image as in \citet{Athanassoula2002} and from the bar axial ratio as in \citet{Aguerri2009}. The $\pm\sigma$ errors on $R_{\rm bar}$ and $S_{\rm bar}$ were estimated by calculating the difference between the adopted value and the highest/lowest measure. The two bars have lengths consistent with the median value found for SB0 galaxies \citep{Aguerri2009} and are both weak according to the classification of \citet{Cuomo2019b}. 

\subsection{Bar pattern speed and rotation rate}
\label{sec:speed_rate}

The integral-field spectroscopy was carried out with the wide field mode of MUSE (Prog. Id. 094.B-0241(A); P.I.: E.M. Corsini) mapping a nominal FOV of $1\,\times\,1$ arcmin$^2$ with a spatial sampling of $0.2$ arcsec pixel$^{-1}$ and covering the wavelength range of $4800-9300$ \AA\ with a spectral sampling of 1.25 \AA\ pixel$^{-1}$ and a nominal spectral resolution of $\rm FWHM=2.51$ \AA. The mean value of the seeing during the observations was $\rm FWHM\sim1.1$ arcsec. The observations were split into different observing blocks which were mosaiced to fully map the galaxies along their photometric major axis for an actual FOV coverage of $1.0\,\times\,1.7$ arcmin$^2$.

\citet{Cuomo2019} and \citet{Buttitta2022} measured the stellar and ionised-gas kinematics of the two galaxies by using the {\sc ppxf} \citep{Cappellari2004} and {\sc gandalf} \citep{Sarzi2006} codes. The spaxels in the datacube were spatially binned with the Voronoi tessellation algorithm \citep{Cappellari2003} to ensure a target signal-to-noise ratio $S/N = 40$ per bin. In each bin, the galaxy spectrum was fitted by convolving the spectra extracted from the ELODIE stellar library ($\sigma_{\rm instr}=13$ km s$^{-1}$, \citealt{Prugniel2001}) with a line-of-sight velocity distribution (LOSVD) modelled with a truncated Gauss-Hermite series \citep{Marel1993, Gerhard1993} in the wavelength range $4800-5600$~\AA. The circular velocity $V_{\rm circ}$ was derived by correcting the stellar streaming motion for the asymmetric drift \citep{Binney1987}.

The pattern speed $\Omega_{\rm bar}$ of both bars was obtained by applying the Tremaine-Weinberg method \citep{TW1984} on the reconstructed image and stellar velocity field of the host galaxies. The value of $\Omega_{\rm bar}$ is given by $\langle V \rangle = \langle X \rangle \sin{(i_{\rm disc})} \Omega_{\rm bar}$. It depends on the disc inclination and on the luminosity-weighted position $\langle X \rangle$ and LOS velocity $\langle V \rangle$ of the stellar component within apertures parallel to the disc major axis and crossing the bar. Finally, the corotation radius $R_{\rm cor}$ and the rotation rate $\mathcal{R}$ values were estimated calculating $R_{\rm cor}$=$V_{\rm circ}/\Omega_{\rm bar}$ and $\mathcal{R}=R_{\rm cor}/a_{\rm bar}$, respectively. The errors on $R_{\rm cor}$ and $\mathcal{R}$ were estimated by using Monte Carlo simulations. We generated a distribution of $R_{\rm cor}$ and $\mathcal{R}$ by accounting for the errors on $R_{\rm bar}$, $i_{\rm disc}$, and $V_{\rm circ}$. The adopted $\pm\sigma$ errors for $R_{\rm cor}$ and $\mathcal{R}$ are calculated as the 16th and 84th percentiles of the distributions.

The two galaxies have similar properties in terms of bar size and strength, but not in terms of bar pattern speed: NGC\,4264 hosts a fast bar ($\mathcal{R}=0.9\pm0.2$) while the bar in NGC\,4277 is slow ($\mathcal{R}=1.8^{+0.5}_{-0.3}$). The different bar rotation rates could be due to a different bar formation mechanism and/or a different DM content in the bar region. 

Although NGC\,4264 possibly forms an interacting pair with NGC\,4261, which is located at a projected distance of 3.5\,arcmin \citep{Schmitt2001}, which corresponds to a quite large physical distance of 4.9\,Mpc, it lacks a strongly disturbed morphology. According to \citet{Cuomo2019}, this suggests that the interaction is weak and not responsible for have triggered the bar formation in NGC\,4264. NGC\,4277 is likely paired with NGC\,4273, which is located at a projected distance of 1.9\,arcmin \citep{vanDriel2000} corresponding to a physical distance of 2.5\,Mpc. \citet{Buttitta2022} argued that the bar formation in NGC\,4277 could have been triggered by their tidal interaction or alternatively, the bar could have been braked by the dynamical friction of a dense DM halo \citep{Debattista2006, Athanassoula2013}. 

\section{Stellar dynamical model}
\label{sec:method}

\subsection{Jeans dynamical models}

Reconstructing the mass distribution of a disc galaxy using unresolved stars as tracers of the gravitational potential is a challenging task due to the non-uniqueness of the light deprojection \citep{Rybicki1987, Gerhard1996}. The bar introduces a further complication since its characterisation requires two additional parameters: the orientation and figure rotation \citep{Lablanche2012}. This increases the degeneracy between the model parameters with different combinations able to reproduce the observed photometric and kinematic properties of the galaxy.

Several methods have been developed to recover the dynamical structure of a galaxy from broad-band imaging and long-slit/integral-field spectroscopy, but the dynamical modelling of barred galaxies is still at an early stage. The recently developed orbit-superposition Schwarzschild models by \citet{Vasiliev2020} and \citet{Tahmasebzadeh2022}, which to date have been applied only to data from N-body simulations, considered the bar pattern speed. \cite{Portail2017} built a dynamical model of the Milky Way to recover its bar pattern speed by using the made-to-measure method as implemented by \cite{deLorenzi2007}.

In general, barred galaxies have been modelled with axisymmetric dynamical models, including the Jeans Anisotropic Modelling \citep[JAM,][]{Cappellari2008, Cappellari2020}, which has been extensively applied to spectroscopic surveys of nearby lenticular and spiral galaxies \citep[e.g.,][]{Williams2009, Cappellari2013, Guo2019}. It models the LOS second velocity moment for galaxies with an axisymmetric mass distribution, including a DM halo, to be compared with the root-mean-square velocity $V_{\rm rms}$ derived from the observed velocity $V_{\rm los}$ and velocity dispersion $\sigma_{\rm los}$. JAM requires the surface-brightness distribution of the galaxy to be described through a Multi-Gaussian Expansion \citep[MGE,][]{MGEPYTHON} parameterisation, which simplifies the solution of Jeans equations to recover the galaxy inclination $i$, mass-to-light ratio $(M/L)_\ast$ of the matter following the light (which may include DM as well as stars), and anisotropy parameter $\beta_z = 1 - \sigma^2_z/\sigma^2_R$, where $\sigma_R$ and $\sigma_z$ are the radial and vertical components of the velocity dispersion, respectively, in a cylindrical coordinate system with the origin in the centre of the galaxy and symmetry axis aligned with its rotation axis. 

\cite{Cappellari2008} compared the JAM and orbit-superposition Schwarzschild models of six fast-rotating lenticular galaxies from the SAURON survey \citep{SAURON}. They have HST and ground-based photometry and SAURON integral-field spectroscopy. The values of $\beta_z$ from JAM are consistent within the uncertainties with those obtained with the Schwarzschild modelling. Since fast rotators show a slightly positive value of $\beta_z$, the inclination-anisotropy degeneracy was removed, constraining $\beta_z>0$. Although there is a small dependence on the anisotropy parameter, overall the $(M/L)_\ast$ values and mass models obtained with the two approaches are also in agreement. This means that the JAM model, with simple and well-motivated assumptions, provides a reasonable description of the mass distribution of lenticular galaxies.

\cite{Cappellari2013} applied the JAM algorithm to 260 nearby early-type galaxies of the ATLAS$^{\rm 3D}$ survey \citep{Cappellari2011}, whose surface-brightness distribution was measured from SDSS and Isaac Newton Telescope imaging and the stellar kinematics were traced out to roughly one effective radius $R_{\rm e}$ with SAURON integral-field spectroscopy. This volume-limited sample was composed of galaxies with a distance $D<42$\,Mpc, absolute magnitude $M_K < -21.5$\,mag, and stellar mass $M_\ast \gtrsim 6 \times 10^9$\,M$_\odot$. The stellar kinematics of most of the sample galaxies are reasonably well reproduced by mass-follows-light models, suggesting that early-type galaxies have a simple internal structure within $1 R_{\rm e}$, and that the DM halo is not dominant. By adding the contribution of a NFW \citep{NFWDM1995} DM halo, \cite{Cappellari2013} estimated a median DM fraction within the effective radius of $f_{\rm DM}(r<R_{\rm e})=0.13$. About one-third of the sample galaxies host a bar and for some of them, the quality of the fit was poor due to the low $S/N$ and/or twisted stellar kinematics or the presence of a strong bar. The recovered values of $(M/L)_\ast$ for the whole sample have an accuracy of 6 per cent, which falls to 15 per cent for the barred galaxies.

\cite{Lablanche2012} analysed realistic simulations of two SB0 galaxies to explore the reliability of the JAM approach in recovering the dynamical parameters of a barred galaxy. The simulated galaxies mimicked NGC\,4442 and NGC\,4754 and were projected at different disc inclinations ($i_{\rm disc} = 25^\circ, 45^\circ, 60^\circ$, and $87^\circ$) and with different bar orientation ($|{\rm PA_{bar}}-{\rm PA_{disc}}|=18^\circ, 45^\circ, 60^\circ$, and $87^\circ$). In general, $i_{\rm disc}$ can be recovered with JAM although this result is biased by the non-uniqueness of the mass deprojection in nearly face-on or edge-on barred systems (with a maximum error $\Delta i_{\rm disc}\sim6^\circ$ for edge-on systems). The value of $\beta_z$ is not well recovered for any disc inclination and bar orientation, because the bar produces a deprojected mass density which is flatter or rounder with respect to the azimuthally averaged one when the bar is viewed side-on or end-on, respectively. This issue is not unique to JAM, but is expected to affect also the axisymmetric dynamical models based on orbit superposition. The recovered $M/L$ depends on the disc inclination and bar orientation. The uncertainty is smaller than 1.5 per cent for systems with $i_{\rm disc}\geq45^\circ$ and $|{\rm PA_{bar}}-{\rm PA_{disc}}|=60^\circ$ and it never exceeds 3 per cent for the other inclinations. The recovered $(M/L)_\ast$ is underestimated (overestimated) if PA$_{\rm disc}<45^\circ$ (PA$_{\rm disc}>45^\circ$). In their tests, they found that the maximum systematic error of 15 per cent on $(M/L)_{\ast}$ occurs when the bar is seen nearly end-on (PA$_{\rm bar}=18^\circ$) or side-on (PA$_{\rm bar}=87^\circ$). They also investigated how the size of the FOV affects the recovered parameters, and concluded that if the FOV extends out to the bar radius, the systematic uncertainty on $(M/L)_{\ast}$ decreases and tends to a limiting value of 10 per cent. However, the mass models of \cite{Lablanche2012} did not include DM halos.

We adopted the JAM method to recover the mass distribution of NGC\,4264 and NGC\,4277, since we are confident that applying such an axisymmetric dynamical model gives a reliable estimate of $(M/L)_\ast$ and DM fraction even in barred galaxies. Both objects are ideal targets according to \cite{Lablanche2012}, because they have an intermediate inclination ($i_{\rm disc}\sim40^\circ$), are not substantially affected by dust, and host a weak bar with an intermediate orientation with respect to the disc major and minor axis ($|{\rm PA_{bar}}-{\rm PA_{disc}}|\sim50^\circ$). According to \cite{Lablanche2012}, in this configuration, we expect to systematically overestimate the $(M/L)_\ast$ by a factor of 10 per cent. This translates into a larger overestimate of the DM fraction in galaxies with a larger content of luminous matter. In addition, the fine spatial sampling, wide FOV, and high spectral resolution of the MUSE integral-field spectrograph made it possible to accurately map the stellar kinematics throughout the galaxy disc. The kinematic maps do not show strong perturbed features as prescribed to minimise the biases in the estimation of the dynamical parameters. Finally, we notice that NGC\,4264 and NGC\,4277 have similar luminosities to those of NGC\,4442 and NGC\,4754. This gives us confidence in extending the findings of \cite{Lablanche2012} to our galaxies.

\subsection{Multi-Gaussian expansion}

\begin{figure*}
    \centering
    \includegraphics[scale=0.38]{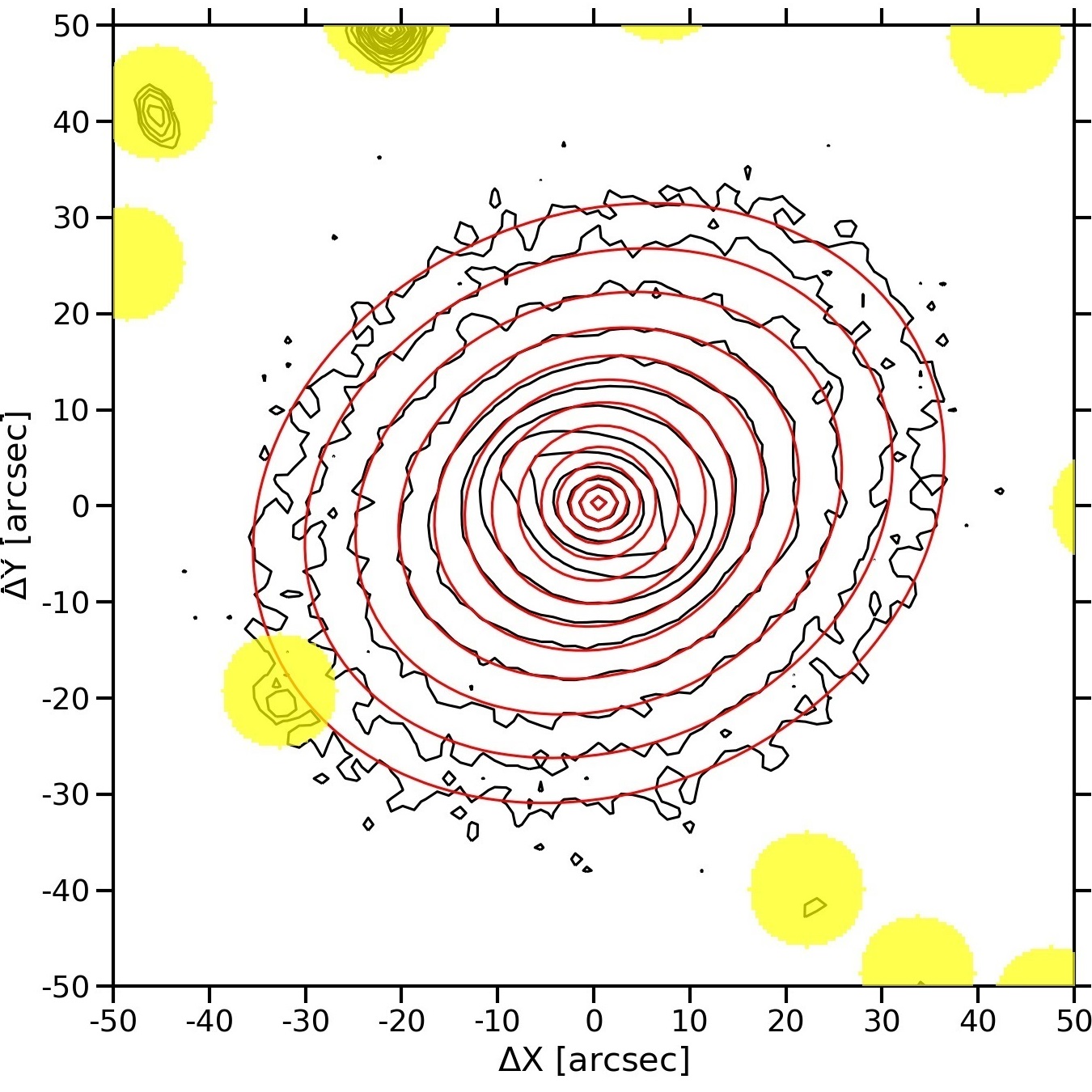}
    \includegraphics[scale=0.38]{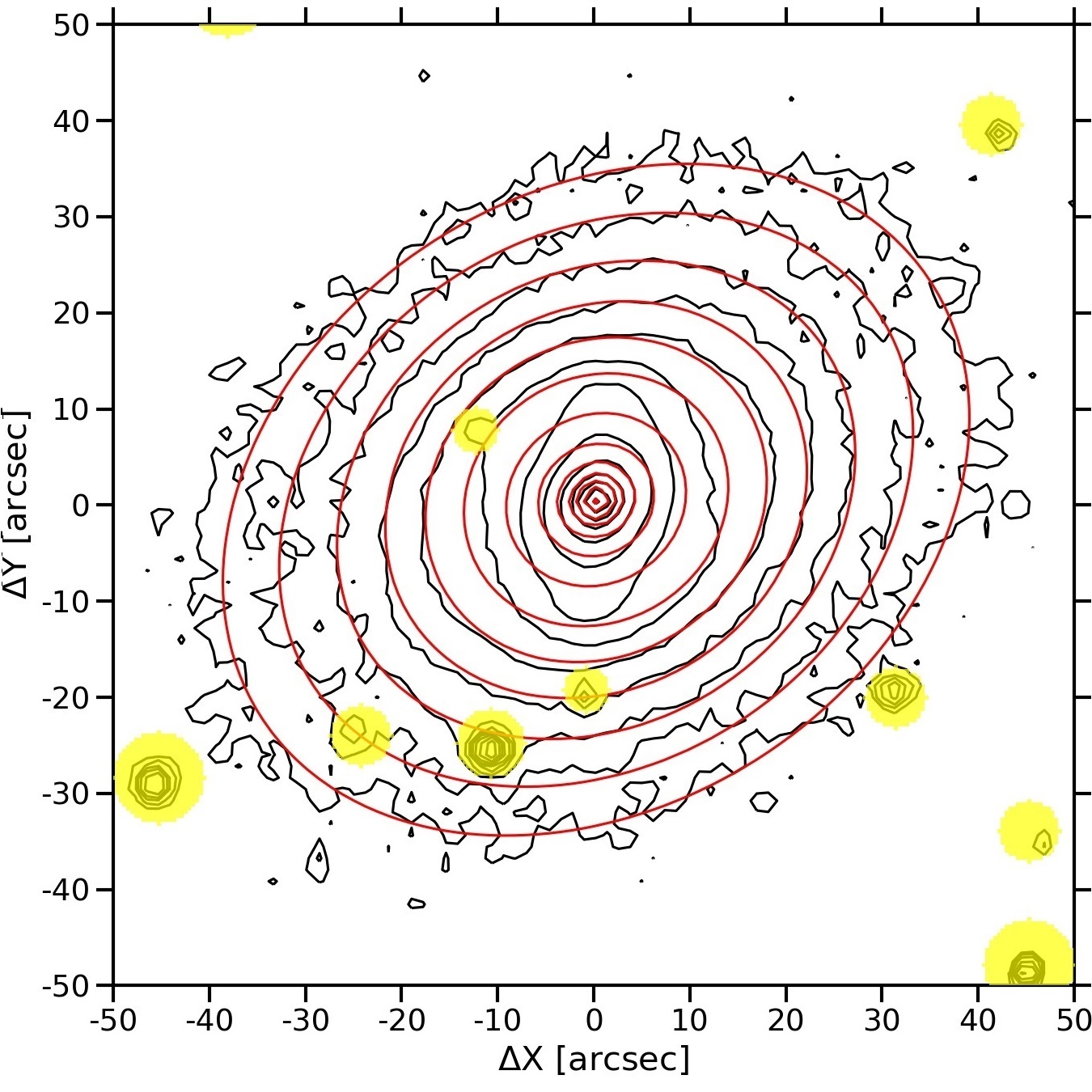}
    \caption{Some reference isophotes for the SDSS $i$-band image (black lines) and MGE model (red lines) of NGC\,4264 (left panel) and NGC\,4277 (right panel). The FOV is $1.7 \times 1.7$ arcmin$^2$ and oriented with North up and East left. Flux levels are normalised to the central surface brightness of the image and the contours are spaced by 0.5 mag arcsec$^{-2}$. While the MGE model was constrained using the original image, the image shown here is binned by $3\times3$ pixels$^2$ to reduce the noise for comparison purposes only. The yellow circles correspond to masked regions.}
    \label{fig:MGE}
\end{figure*}

To obtain a model for the luminosity volume density of both NGC\,4264 and NGC\,4277, we started by parameterising the $i$-band surface brightness of the sky-subtracted image of each galaxy as the sum of a set of Gaussian components by using the MGE Python code, which is based on \citet{MGEPYTHON}. The MGE method allows for a simple reconstruction of the intrinsic surface brightness distribution provided that the point spread function (PSF) can be approximated as a sum of Gaussian components. The intrinsic surface brightness distribution is then easily deprojected into the luminosity volume density, which is also parameterised as the sum of a set of Gaussian components.

We adopted the centre coordinates derived for the two galaxies by \citet{Cuomo2019} and \citet{Buttitta2022}. We estimated the root mean square of the sky surface brightness by performing a statistical analysis on different regions of the images containing exclusively the sky contribution. These areas were selected in empty regions, which were free of objects and far from the target galaxy to avoid the contamination of the light of field stars and background galaxies, as well as of the galaxy itself. Finally, we modelled the PSF by applying the MGE algorithm on a bright, isolated, and round-shaped field star constraining the best-fitting Gaussians to have a perfect circular shape. The surface brightness distribution of NGC\,4264 is characterised by an isophotal twist in the outermost regions. The internal (PA$_{\rm in}=114\fdg0\pm1\fdg2$) and external regions (PA$_{\rm out}=122\fdg8\pm2\fdg4$) of the disc have different orientations but the same shape ($\epsilon=0.20\pm0.02$) as found by \citet{Cuomo2019}. They argued that the isophotal outer twist is not representative of the actual orientation of the disc. We decided to constrain the Gaussians with the geometric parameters of the internal disc which is mapped by the measured stellar kinematics. The radial profiles of PA and $\epsilon$ derived by \citet{Buttitta2022} from the surface brightness distribution of NGC\,4277 show a clear disc-dominated region with a well-defined orientation (PA$ = 123\fdg3\pm0\fdg3$) and shape ($\epsilon=0.24\pm0.02$).

We obtained the MGE best-fitting model to the galaxy surface brightness by keeping constant the centre and position angle of the Gaussians derived by \citet{Cuomo2019} and \citet{Buttitta2022}, while further restricting the range of the resulting axial ratios of the Gaussian components to [$q_{\rm min}$, 1] where $q_{\rm min}=1-\epsilon_{\rm disc}$. This ensured that the permitted galaxy inclinations were not limited to a narrower range than that allowed by the data \citep[e.g.,][]{Scott2013, Pagotto2019}.

We show a few representative isophotes of the $i$-band images of NGC\,4264 and NGC\,4277 and compare these to the corresponding MGE best-fitting contours in Fig.~\ref{fig:MGE}. The MGE algorithm provided the total luminosity in counts, root mean square in pixels, and axial ratio for each best-fitting Gaussian parameterising the intrinsic surface brightness distribution. We converted the output parameters into physical units by using the flux calibration and spatial scale of the images given by \citet{Cuomo2019} and \citet{Buttitta2022} and correcting for cosmological dimming, $K$-correction \citep{Chilingarian2012}, and Galactic extinction \citep{Schlafly2011}. We adopted $M_{i,\odot}=4.53$ mag as the absolute magnitude of the Sun in the SDSS $i$-band \citep{Willmer2018}. We list the MGE best-fitting parameters of the intrinsic surface brightness distribution of NGC\,4264 and NGC\,4277 in Table~\ref{tab:MGE}.

\begin{table}
\centering
\renewcommand{\tabcolsep}{0.3cm}
\renewcommand{\arraystretch}{1.4}
\begin{tabular}{c c c c c c }
\hline
\multicolumn{3}{c}{NGC\,4264}  & \multicolumn{3}{c}{NGC\,4277} \\
\hline
\hline
$I_0$ & $\sigma$ & $q$  & $I_0$ & $\sigma$ & $q$   \\
$[{\rm L}_\odot \ {\rm pc}^{-2}]$ & $[ \rm kpc]$  & & $[{\rm L}_\odot \ {\rm pc}^{-2}]$ & $[ \rm kpc]$ &  \\
(1) & (2) & (3) & (1) & (2) & (3) \\
\hline
  15427.0  & 0.04   & 0.80 & 5106.5  & 0.08  & 0.88   \\
  4309.5   & 0.13   & 0.80 & 1023.8  & 0.20  & 0.85   \\
  691.4    & 0.34   & 0.80 & 353.5   & 0.48  & 0.92   \\
  1141.1   & 0.43   & 1.00 & 158.6   & 1.52  & 0.91   \\
  443.8    & 1.01   & 1.00 & 41.9    & 3.78  & 0.76   \\
  219.3    & 1.72   & 0.80 &         &       &        \\
  64.8     & 3.86   & 0.81 &         &       &        \\
\hline
\end{tabular}
\caption{Best-fitting parameters of the Gaussian components in the MGE model of the $i$-band surface brightness distribution of NGC\,4264 and NGC\,4277. (1): Central luminosity surface density. (2) Standard deviation. (3) Axial ratio.}
\label{tab:MGE}
\end{table}

\subsection{Axisymmetric Jeans anisotropic model}

With the MGE models at hand, we proceeded to use the JAM Python code based on \citet{Cappellari2008} to build Jeans axisymmetric dynamical models for NGC\,4264 and NGC\,4277 in order to derive the DM content within the bar region.

We selected the best-fitting JAM model by minimising the $\chi^2$ difference between the predicted second moment of the velocity field and MUSE stellar kinematics measured by \citet{Cuomo2019} and \citet{Buttitta2022}. To this aim, we obtained, for each spatial bin, the observed $V_{\rm rms}$ as

\begin{equation}
\nonumber
V_{\rm rms} \equiv \sqrt{ V_{\rm los}^2 + \sigma_{\rm los}^2}
\end{equation}

\noindent and its corresponding error 

\begin{equation}
\nonumber
\epsilon_{V_{\rm rms}} \equiv \frac{1}{V_{\rm rms}} \sqrt{(V_{\rm los}\epsilon_{V_{\rm los}})^2 + (\sigma_{\rm los}\epsilon_{\sigma_{\rm los}})^2}.
\end{equation}

\noindent We discarded the bins with values of $V_{\rm los}$ and $\sigma_{\rm los}$ with relative uncertainties $\epsilon_{V_{\rm los}} / V_{\rm los}>1$ and $\epsilon_{\sigma_{\rm los}}/\sigma_{\rm los}>1$. In both galaxies, the maximum uncertainty on $V_{\rm rms}$ never exceeds $\epsilon_{V_{\rm rms}}\sim8$\,km s$^{-1}$. Finally, we symmetrised the resulting values with respect to the disc major axis by means of the Python algorithm {\sc plotbin}. We show the $V_{\rm rms}$ maps of NGC\,4264 and NGC\,4277 in the top-left panels of Fig.~\ref{fig:NGC4264total} and \ref{fig:NGC4277total}, respectively. 

We modelled the total mass distribution of both galaxies as the sum of two components

\begin{equation}
\nonumber
\rho = \rho_\ast + \rho_{\rm halo}
\end{equation}

\noindent where $\rho_\ast$ is the mass volume density of matter (either luminous or dark) that is distributed like the stars and $\rho_{\rm halo}$ is the mass volume density of DM distributed in a spherical halo. We built a set of mass-follows-light models by assuming that the mass volume density $\rho_\ast$ follows the luminosity volume density $\nu_\star$ derived by deprojecting the intrinsic surface brightness distribution obtained from the MGE

\begin{equation}
\nonumber
\rho_\ast = (M/L)_\ast \nu_\ast
\end{equation}

\noindent These models have three free parameters that are optimised while matching the observed $V_{\rm rms}$. They are $i_{\rm disc}$, $(M/L)_\ast$, and $\beta_z$. We adopted radially constant values for both $(M/L)_\ast$ and $\beta_z$. 

Then, we included the contribution of the DM halo $\rho_{\rm halo}$, for which we considered

\begin{enumerate}

\item a quasi-isothermal \citep[QI,][]{Binney1987} radial profile of the mass volume density

\begin{equation}
\rho_{\rm QI}(r) = \frac{\rho_0}{ 1 + \Bigl( \frac{r}{r_{\rm c}}\Bigr )^2}, 
\nonumber
\end{equation}
    
\noindent where $\rho_0$ and $r_{\rm c}$ are the central mass volume density and core radius, respectively. 

\item a NFW radial profile of the mass volume density

\begin{equation}
    \rho_{\rm NFW}(r) = \frac{\rho_{\rm s}}{\frac{r}{r_{\rm s}}\Bigl (1 + \frac{r}{r_{\rm s}}\Bigr )^2 }, 
    \nonumber
\end{equation}
    
\noindent where $\rho_{\rm s}$ and $r_{\rm s}$ are the scale mass volume density and scale radius, respectively. We reduced the number of free parameters by adopting the following parameterisation

\begin{equation}
    \rho_{\rm NFW}(r) = \frac{M_{\rm vir}}{4\pi A(c_{\rm vir})}\cdot \frac{1}{r(r_{\rm s} + r)}, 
    \nonumber
\end{equation}

\noindent where the virial mass $M_{\rm vir}$ and coefficient $A_{\rm c_{\rm vir}}$ are respectively given by 

\begin{equation}
    \nonumber
    M_{\rm vir} = \frac{4\pi}{3}r^3_{\rm vir}\,\rho_{\rm crit}\Omega_{\rm M}\Delta_{\rm vir},
\end{equation}

\noindent with $\rho_{\rm crit} = 1.37 \cdot 10^{-7}$ M$_\odot$\,pc$^{-3}$, $\Omega_{\rm M}=0.27$ and $\Delta_{\rm vir}=200$, and 

\begin{equation}
    A(c_{\rm vir}) = \log{(1 + c_{\rm vir})}-\frac{c_{\rm vir}}{1+c_{\rm vir}},
    \nonumber
\end{equation}

\noindent where $c_{\rm vir}=r_{\rm vir}/r_{\rm s}$ is the concentration parameter. We followed the $M_{\rm vir} - c_{\rm vir}$ relation by \cite{Klypin2011} to derive

\begin{equation}
    c_{\rm vir} = 9.6 \Bigl ( \frac{0.7 M_{\rm vir}}{10^{12}}\Bigr )^{-0.075}.
    \nonumber
\end{equation}

\noindent In this way, the mass model has only a free parameter $M_{\rm vir}$.

\item a generalised NFW \citep[gNFW,][]{Barnabe2012} radial profile of the mass volume density
    
\begin{equation}
    \rho_{\rm gNFW}(r) = \rho_{\rm s} \Bigl ( \frac{r}{r_{\rm s}} \Bigr ) ^\gamma \cdot \Bigl (\frac{1}{2} + \frac{1}{2}\frac{r}{r_{\rm s}} \Bigr )^{- (\gamma + 3)},
    \nonumber
    \end{equation}

\noindent where $\rho_{\rm s}$ and ${r_{\rm s}}$ are the scale mass volume density and scale radius, respectively, while the $\gamma$ parameter gives the inner slope of the radial profile. We constrained it in the range $-2 < \gamma < 0$, with $\gamma=0$ corresponding to a cored profile and $\gamma=-1$ to a NFW profile. 

\end{enumerate}

For each mass model, we computed the radial profiles of the enclosed mass and circular velocity for the stars, DM in the halo, and their sum. We used Monte Carlo simulations to estimate the $1\sigma$ confidence intervals. At each radius, we generated a distribution of enclosed mass and circular velocity taking into account the errors on the best-fitting parameters of the mass model. The adopted $\pm\sigma$ errors of enclosed mass and circular velocity for the stars, DM in the halo, and their sum are calculated as the 16th and 84th percentiles of the distributions.

For the three mass models with a DM halo, we calculated the fraction of DM within the bar region as 

\begin{equation}
\nonumber
f_{\rm DM, bar} = \frac{M_{\rm DM}(r<a_{\rm bar})}{M_{\rm DM}(r<a_{\rm bar}) + M_{\rm \star}(r<a_{\rm bar})}.
\end{equation}

\noindent We calculated the corresponding $\pm\sigma$ errors as the 16th and 84th percentiles of the distribution of $f_{\rm DM, bar}$ that we built from the same Monte Carlo simulations previously generated.

The JAM code allows the inclusion of the contribution of a central supermassive black hole (SMBH), which is modelled through the MGE formalism as a Gaussian having mass $M_\bullet$, axial ratio $q=1$ and $3\sigma \lesssim r_{\rm min}$, with $r_{\rm min}$ defined as the smallest distance from the SMBH that can be chosen $r_{\rm min} \sim \sigma_{\rm PSF}$ \citep{Cappellari2008, Cappellari2020}. Due to the limited spatial resolution of the available ground-based kinematic observations, we can not constrain the mass $M_\bullet$ of the central SMBH. Therefore, we decided to adopt the SMBH mass value given by the $M_\bullet - \sigma_{\rm e}$ relation \citep{Kormendy2013}. We estimated the bulge effective velocity dispersion from the MUSE kinematic map and the reconstructed image as the luminosity-weighted average of the observed $V_{\rm rms}$ inside an elliptical region with a semi-major axis equal to half of the bulge effective radius $R_{\rm e, bulge}$ and the same axial ratio $q_{\rm bulge}$ of the bulge. We adopted the values obtained by \cite{Cuomo2019} and \cite{Buttitta2022} with a parametric photometric decomposition of the SDSS images of NGC\,4264 ($R_{\rm e, bulge}=1.53$\,arcsec, $q_{\rm bulge}=0.77$) and NGC\,4277 ($R_{\rm e, bulge}=1.77$\,arcsec, $q_{\rm bulge}=0.84$).

\section{Results}
\label{sec:results}

For both galaxies, we obtained the best-fitting parameters for all the mass models without and with a DM halo. We analysed the mass-follows-light models as well as the models with a QI, NFW, and gNFW DM halo to constrain the DM fraction within the bar region of NGC\,4264 and NGC\,4277.  

\subsection{NGC~4264}

\begin{figure*}
\centering
\includegraphics[scale=0.34]{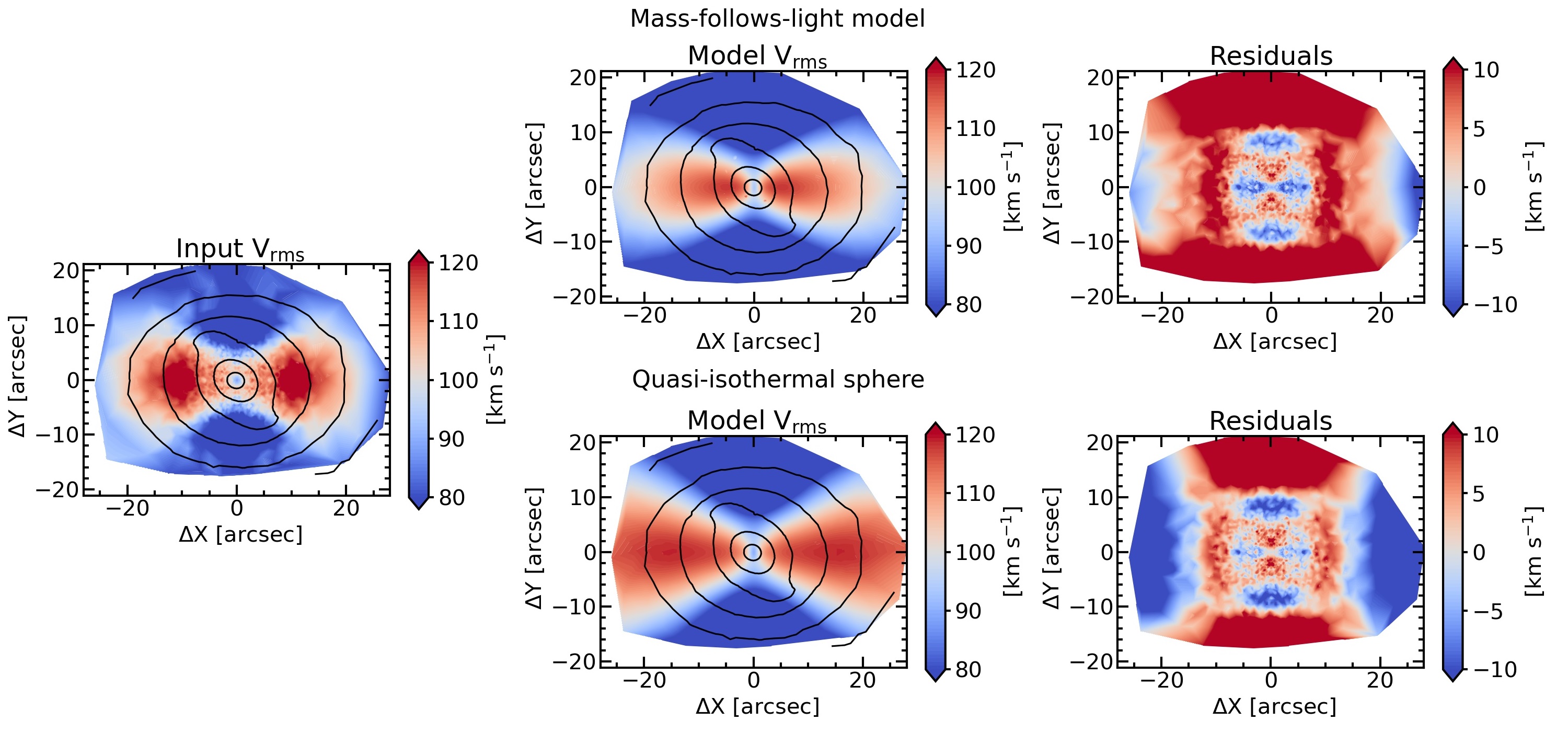}
\includegraphics[scale=0.38]{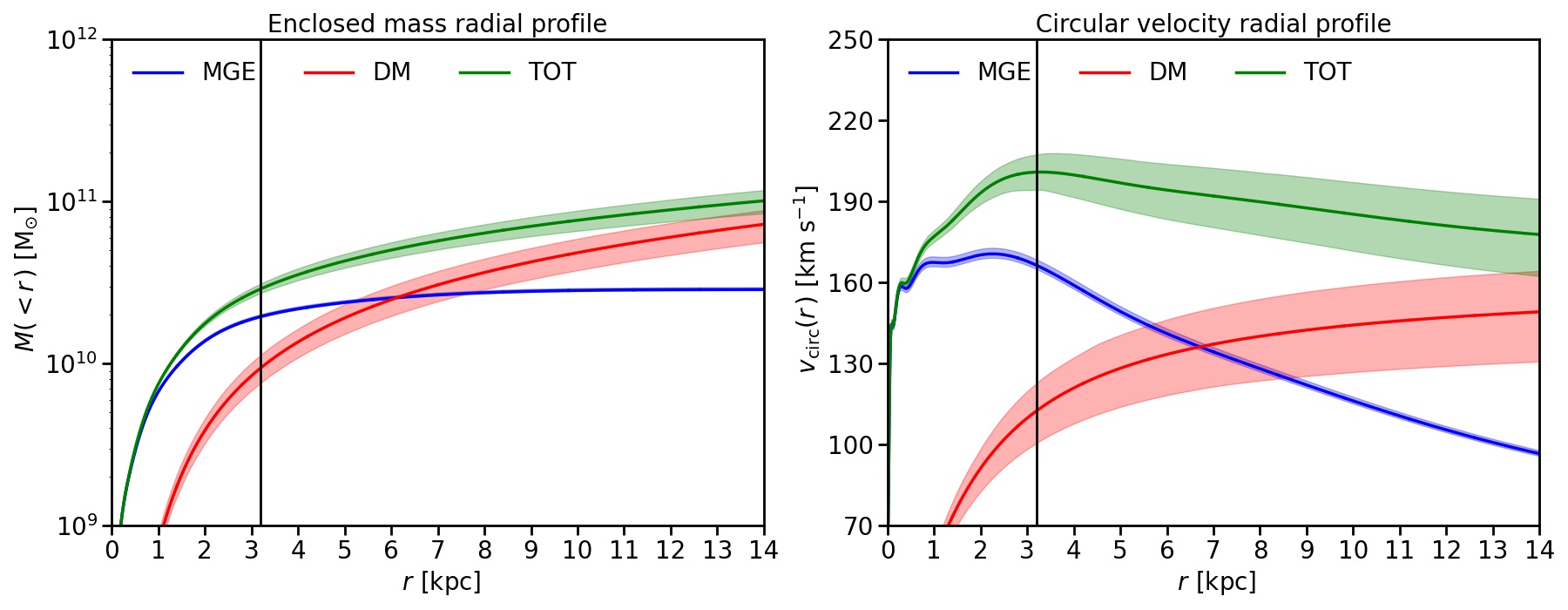}
\caption{Top panels: Symmetrised $V_{\rm rms}$ from the MUSE stellar kinematics of NGC\,4264 (left panel), bisymmetric second velocity moment predicted by the mass-follows-light model (top-centre panel) and by the mass model with a QI halo (bottom-centre panel), and residuals of the observed and modelled $V_{\rm rms}$ (top-right and bottom-right panels). Bottom panels: Radial profiles of enclosed mass (left panel) and circular velocity (right panel) of NGC\,4264 for the mass model with a QI halo. The contributions of the stars (blue line) and DM (red line) are plotted with their sum (green line). The vertical solid and dashed lines mark the bar radius and the extension of kinematic data, respectively. The shaded areas represent the $\pm\sigma$ errors calculated from Monte Carlo simulations.} 
\label{fig:NGC4264total}
\end{figure*}

We construct a starting set of mass models by considering the galaxy inclination as a free parameter. The mass models with a DM halo return the same value (QI, gNFW) or a value consistent within errors (NFW) with that obtained by \citet{Cuomo2019} from the isophotal analysis of NGC\,4264 ($i_{\rm disc} =  36\fdg7$), whereas the mass-follows-light model gives a much larger value ($i =  42\fdg9\pm1\fdg3$). 

We repeat the analysis after masking the kinematic bins within a circular region of radius $r=8$ arcsec, which corresponds to the bar radius projected onto the sky plane. In this way, we consider only the kinematic data measured in the disc region. But, all the mass models converge to an edge-on solution with $\beta_z\sim-1.45$ as a consequence of the inclination-anisotropy degeneracy \citep{Krajnovic2005, Lablanche2012}. We verify that this result is not driven by the twisted structure of the disc of NGC\,4264. The kinematic bins in the external disc ($r>25$ arcsec) account for less than 1 per cent of the data and the mass models based on a different MGE decomposition of the SDSS image, where we mask the external disc, do not improve the fit. We perform a further sanity check by masking the kinematic bins in the central circular region of radius $r=2.5$ arcsec to minimise biases due to an uncorrected estimation of the PSF and/or the SMBH mass. Then, we decide to fix the inclination to build the final set of mass models. We adopt $i=36\fdg5$ which corresponds to the photometric value of \citet{Cuomo2019}. An intrinsic flattening $q_0=0.05$ is adopted in JAM modelling to derive $i$ from the observed axial ratio, whereas $i_{\rm disc}$ is obtained by \citet{Cuomo2019} assuming an infinitesimally thin disc. The choice of a fixed inclination allows a straightforward comparison between the best-fitting parameters of the different mass models, which we list in Table~\ref{tab:JAM_results} together with their reduced $\chi^2$.

We find a slightly larger mass-to-light ratio for the mass-follows-light model ($(M/L)_{\ast,i}\sim2.6$ M$_\odot$\,L$_\odot^{-1}$) with respect to the mass models with a DM halo ($(M/L)_{\ast,i}\sim2.1$ M$_\odot$\,L$_\odot^{-1}$). This is expected if there is a small amount of DM, which is not distributed like the stars. Since we are interested in recovering the mass distribution of NGC\,4264 and not in its orbital structure, we were not concerned by the fact that the mass-follows-light model was not able to constrain $\beta_z$ and the mass models with the DM halo returned a remarkably negative value of anisotropy ($\beta_z \sim -7.3$).

The observed $V_{\rm rms}$ is characterised by a central local minimum of $\sim 90$ km\,s$^{-1}$ with a remarkable double-peaked structure with a maximum of $\sim 120$ km\,s$^{-1}$ at $|r|\sim15$ arcsec along the galaxy major axis decreasing to $\sim 80$ km\,s$^{-1}$ at $|r|\ga20$ arcsec (Fig.~\ref{fig:NGC4264total}, left panel). Although the overall shape of the iso-velocity contours is reproduced by the mass-follows-light model (Fig.~\ref{fig:NGC4264total}, top-centre panel), it fails to match the location and amplitude of the double peak of $V_{\rm rms}$ (Fig.~\ref{fig:NGC4264total}, top-right panel). The mass models with a DM halo provide a better fit to the observed $V_{\rm rms}$ (Fig.~\ref{fig:NGC4264total}, bottom-centre panel), although they do not correctly reproduce the decrease measured at large radii along the galaxy major axis (Fig.~\ref{fig:NGC4264total}, bottom-right panel). At face value, the mass model with the QI halo is marginally better than those with the NFW ($\Delta \chi^2_\nu = 0.09$) and gNFW halo ($\Delta \chi^2_\nu = 0.01$). 

The DM fraction within the bar of the best-fit model is $f_{\rm DM, bar}^{\rm QI}=0.33\pm0.04$, which is compatible, within 1$\sigma$ uncertainty, with that predicted by the mass models with a NFW ($f_{\rm DM, bar}^{\rm NFW}=0.35\pm0.01$) and gNFW halo ($f_{\rm DM, bar}^{\rm gNFW}=0.34\pm0.05$). This suggests that the mass budget of NGC\,4264 is baryon-dominated in the radial range mapped by the kinematic data. We show in the top panels of Fig.~\ref{fig:NGC4264total} the maps of the second velocity moments predicted of the best-fitting mass models of NGC\,4264 without and with a QI halo to be compared with the map of observed $V_{\rm rms}$. The corresponding radial profile of the enclosed mass and circular velocity for the stars, DM, and their sum are given in the bottom panels of Fig.~\ref{fig:NGC4264total} for the best-fitting mass model with the QI halo. 

The circular velocity profile in the region between the bar end and the edge of the kinematic data ($3 \leq r \leq 5$\,kpc) is characterised by a weak decline. This means that, in this radial range, the DM halo does not play a dominant role with respect to the luminous component. This is a local trend which has been commonly observed in galaxies with massive bulges, with more luminous galaxies having on average more strongly declining rotation curves. At large radii, however, all declining rotation curves flatten out, indicating that substantial amounts of DM must be present in these galaxies too \citep{Noordermeer2007, Kalinova2017, Frosst2022}.

\begin{table*}
\centering
\renewcommand{\tabcolsep}{0.9cm}
\renewcommand{\arraystretch}{1.35}
\begin{tabular}{c c c c c }
\hline
Model               & Parameter    &                              & NGC\,4264       & NGC\,4277      \\
\hline
\hline
mass-follows-light  & $(M/L)_{\ast,i}$        & [M$_\odot$\,L$_\odot^{-1}$]  & $2.55\pm0.02$   & $2.36\pm0.05$  \\
                    & $i$          & $[^\circ]$                   & (36.5)          & (40.6)         \\
	              & $\beta_z$    &                              & unc.            & unc.           \\
                    & $\chi^2_\nu$ &                              & 5.54            & 5.81           \\
\hline
QI       & $(M/L)_{\ast,i}$        & [M$_\odot$\,L$_\odot^{-1}$]  & $2.18\pm0.05$   & $1.72\pm0.02$   \\
	              & log$_{10}(\rho_0/$M$_\odot$\,pc$^{-3}$) &   & $-0.61\pm0.13$  & $-1.04\pm0.04$  \\
	              & $r_{\rm c}$  & [kpc]                        & $1.33\pm0.28$   & $2.46\pm0.24$   \\
	              & $i$          & $[^\circ]$                   & (36.5)          & (40.6)          \\
	              & $\beta_z$    &                              & $-7.12\pm1.73$  & $-1.48\pm0.19$  \\
                    & $\chi^2_\nu$ &                              & 3.98            & 1.54            \\
                    & $f_{\rm DM, bar}$ &                         & $0.33\pm0.04$   & $0.53\pm0.02$   \\
\hline
NFW      & $(M/L)_{\ast,i}$        & [M$_\odot$\,L$_\odot^{-1}$]  & $2.14\pm0.03$   & $1.55\pm0.03$   \\
	              & log$_{10}(M_{200}/{\rm M}_\odot$) &         & $13.94\pm0.12$  & $13.33\pm0.05$  \\
	              & $i$          & $[^\circ]$                   & (36.5)          & (40.6)          \\
	              & $\beta_z$    &                              & $-7.86\pm1.99$  & $-1.55\pm0.20$  \\
                    & $\chi^2_\nu$ &                              & 4.07            & 1.62            \\
                    & $f_{\rm DM, bar}$ &                         & $0.35\pm0.01$   & $0.56\pm0.01$   \\
\hline
gNFW     & $(M/L)_{\ast,i}$        & [M$_\odot$\,L$_\odot^{-1}$]    & $2.14\pm0.05$     & $1.69\pm0.04$   \\
	              & log$_{10}(\rho_{\rm s}/$M$_\odot$\,pc$^{-3}$) & & $-1.20\pm0.15$  & $-2.14\pm0.36$  \\
	              & $r_{\rm s}$  & [kpc]                          & $2.25\pm0.53$     & $11.3\pm6.6$    \\
	              & $\gamma$     &                                & (0)               & $-0.34\pm0.26$  \\
	              & $i$          & $[^\circ]$                     & (36.5)            & (40.6)          \\
	              & $\beta_z$    &                                & $-6.86\pm1.65$    & $-1.49\pm0.19$  \\
                    & $\chi^2_\nu$ &                                & 3.99              & 1.55            \\
                    & $f_{\rm DM, bar}$  &                          & $0.34\pm0.05$     & $0.53\pm0.18$   \\
\hline
\end{tabular}
\caption{Best-fitting parameters of the mass models of NGC\,4264 and NGC\,4277. 
Bracket values refer to fixed parameters, while unconstrained values (having relative errors larger than 1) are labelled.}
\label{tab:JAM_results}
\end{table*}

For the QI model, the contribution of DM starts to dominate the mass budget far beyond the bar region at a galactocentric distance $r\ga7$ kpc). As a further check of our dynamical modelling, we derive the mean circular velocity of the inner disc in the radial range ($3.3 \leq r \leq 4.2$\,kpc) adopted by \cite{Cuomo2019} to estimate the circular velocity by correcting the stellar streaming motion for asymmetric drift. We find $V_{\rm circ}^{\rm QI}=200\pm7$ km\,s$^{-1}$ which is consistent within 2$\sigma$ error with the asymmetric drift estimate $V^{\rm AD}_{\rm circ} = 189 \pm 10$ km s$^{-1}$ by \citet{Cuomo2019}.

We compare our results with those obtained by \citet{Cappellari2013}, who modelled NGC\,4264 using a JAM mass model with a NFW halo. Their stellar kinematics maps were obtained with the SAURON integral-field spectrograph, covering a smaller FOV ($0.55 \times 0.7$ arcmin$^2$) and having a lower angular resolution (FWHM$=1.5$\,arcsec) with respect to ours. Nevertheless, \citet{Cappellari2013} reported that the DM fraction within the galaxy's effective radius ($R_{\rm e} = 13.4$\,arcsec) is $f_{\rm DM}(r<R_{\rm e})=0.31$ with a maximum circular velocity of $V_{\rm circ,max}=191$\,km\,s$^{-1}$. We find a consistent value of $f_{\rm DM}^{\rm NFW}(r~<~R_{\rm e})=0.28 \pm 0.01$, but a larger value of $V_{\rm circ,max}^{\rm NFW}=260 \pm 4$\,km\,s$^{-1}$ for the mass model with a NFW halo. The discrepancy between the two values of the circular velocity could due to the different extensions of the adopted data.

\subsection{NGC~4277}

\begin{figure*}
\centering
\includegraphics[scale=0.34]{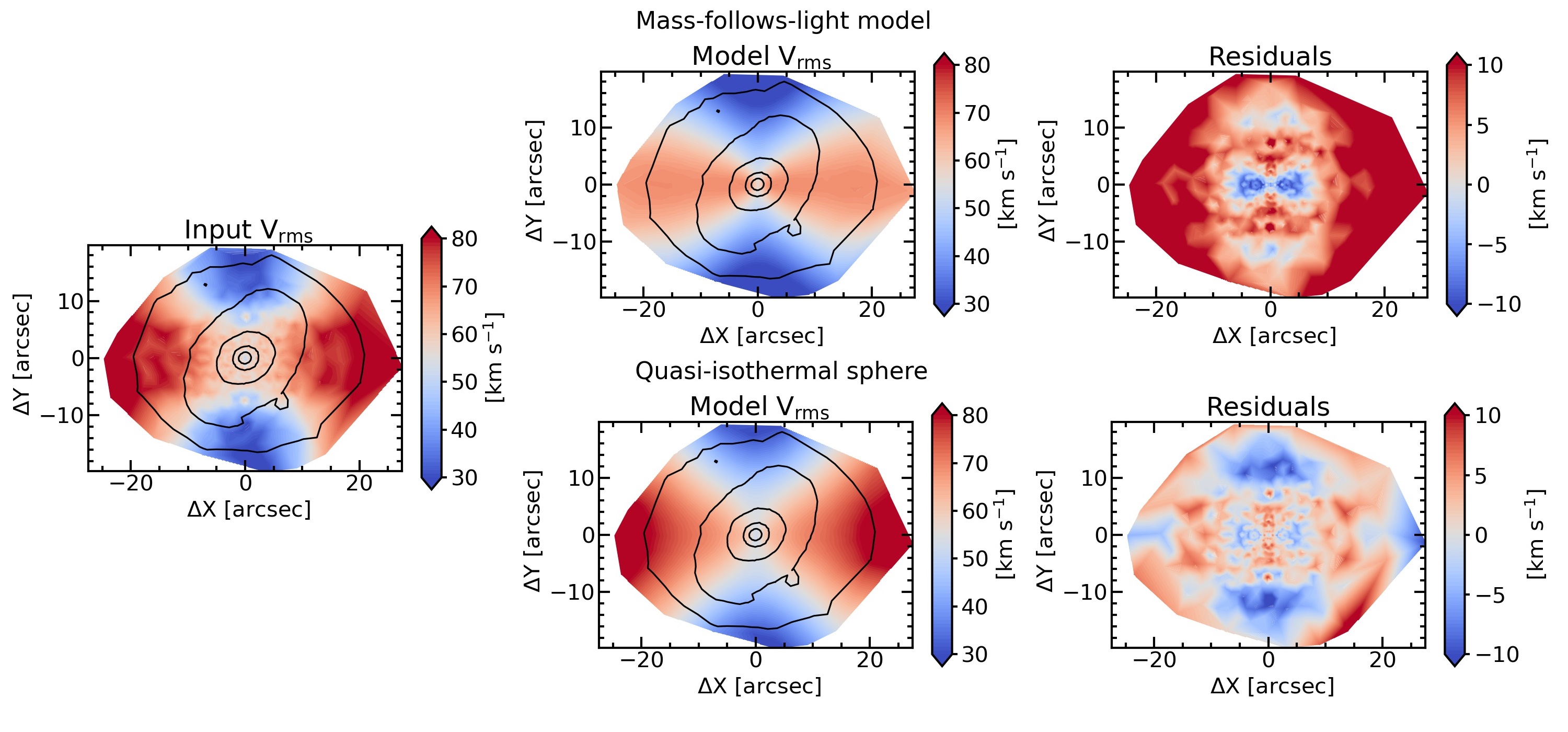}
\includegraphics[scale=0.38]{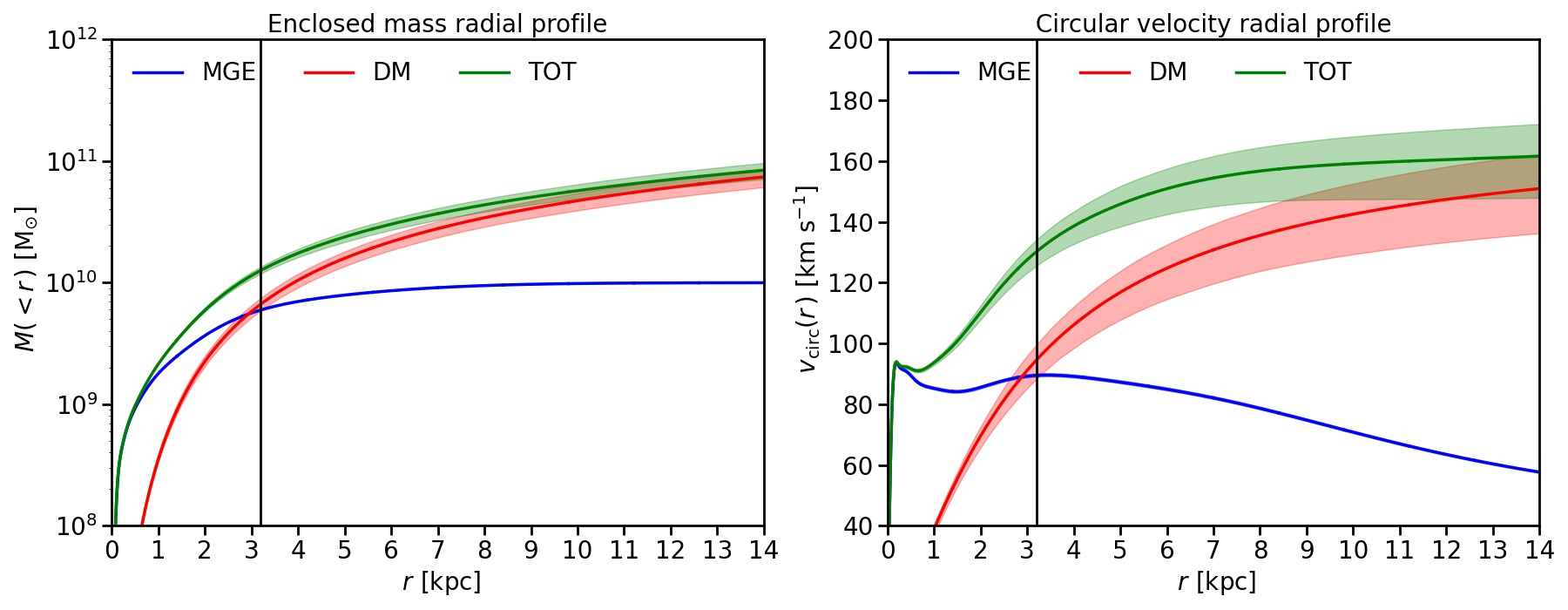}
\caption{Same as Fig.~\ref{fig:NGC4264total}, but for NGC\,4277.} 
\label{fig:NGC4277total}
\end{figure*}

As done for NGC\,4264, we construct a starting set of mass models by considering the galaxy inclination as a free parameter. Considering the intrinsic flattening, the mass models with a DM halo return a consistent value ($i =  40\fdg6$) with that obtained by \citet{Buttitta2022} from the isophotal analysis of NGC\,4277 ($i_{\rm disc} =  40\fdg7$), whereas the mass-follows-light model gave $i =  46\fdg3\pm6\fdg0$. We verify that all the mass models recovered the photometric inclination after masking the kinematic bins belonging to the bar-dominated region ($r<8$ arcsec). In this way, we rely only on the kinematic bins of the disc. We perform a further check by masking the central kinematic bins ($r<2.5$ arcsec) to avoid issues related to the wrong PSF and/or the SMBH mass and we find the same model parameters. Finally, we decide to fix the inclination to build the final set of mass models. We adopt the value consistent with the photometric estimate to allow a straightforward comparison between the best-fitting parameters of the different mass models, which are given in Table~\ref{tab:JAM_results} together with the reduced $\chi^2$ of the mass models.

We find a much larger mass-to-light ratio for the mass-follows-light model ($(M/L)_{\ast,i}\sim2.4$ M$_\odot$\,L$_\odot^{-1}$) with respect to the mass models with a DM halo ($(M/L)_{\ast,i}\sim1.7$ M$_\odot$\,L$_\odot^{-1}$). This is expected if there is DM in addition to the stars. We are not able to constrain $\beta_z$ with the mass-follows-light model, whereas all the mass models with a DM halo returned the same negative value within errors for the anisotropy ($\beta_z\sim-1.5$). 

Although more than 60 per cent of the kinematic bins of NGC\,4277 are located in the bar-dominated region, where the stars are expected to dominate the galaxy mass (Fig.~\ref{fig:NGC4277total}, left panel), the predicted second velocity moment of the mass-follows-light model does not match the observed $V_{\rm rms}$ in terms of amplitude ($V_{\rm rms}\sim80$\,km\,s$^{-1}$ for $r\ga12$ arcsec) and shape of the iso-velocity contours (Fig.~\ref{fig:NGC4277total}, top-centre panel), contrary to the models with a DM halo (Fig.~\ref{fig:NGC4277total}, bottom-centre panel). Therefore, we conclude that the best-fitting mass model of NGC\,4277 requires a DM halo. As for NGC\,4264, the mass model with the QI halo fits slightly better than those with the NFW ($\Delta \chi^2_\nu = 0.08$) and gNFW halo ($\Delta \chi^2_\nu = 0.01$). The DM fraction within the bar is $f_{\rm DM, bar}^{\rm QI}=0.53\pm0.02$ which is fully consistent with the fractions predicted by the mass models with a NFW ($f_{\rm DM, bar}^{\rm NFW}=0.56\pm0.01$) and gNFW DM halo ($f_{\rm DM, bar}^{\rm gNFW}=0.53\pm0.18$). These findings suggest that NGC\,4277 hosts a considerable amount of DM, which is not tied to the stars, within the radial range mapped by the kinematic data. This holds no matter the adopted radial profile of the DM mass volume density. We argue that the large amount of DM in the inner regions of NGC\,4277 is responsible for the slowdown of its bar. 

We show in the top panels of Fig.~\ref{fig:NGC4277total} the maps of the second velocity moments predicted by the best-fitting mass models of NGC\,4277 without and with the QI halo to be compared with the map of observed $V_{\rm rms}$. The corresponding radial profile of the enclosed mass and circular velocity for the stars, DM, and their sum of the best-fitting mass model with the QI halo are shown in the bottom panels of Fig.~\ref{fig:NGC4277total}.  

The contribution of the DM starts to dominate the mass budget just outside the bar region ($r\ga3$ kpc) and the circular velocity flattens out at a larger galactocentric distance ($r\ga5$ kpc) as expected for a DM-dominated region. We derive the mean value $V_{\rm circ}^{\rm QI}=136\pm4$ km\,s$^{-1}$ of the circular velocity in the same radial range ($2.1\leq r \leq 5.9$\,kpc) adopted by \cite{Buttitta2022} to estimate the circular velocity by correcting the stellar streaming motion for asymmetric drift. They found $V_{\rm circ}^{\rm AD} =148\pm5$ km\,s$^{-1}$. The two values are consistent with each other within $2\sigma$ errors.

\section{Conclusions}
\label{sec:conclusion}

We have built Jeans axisymmetric dynamical models for the two barred lenticular galaxies NGC\,4264 and NGC\,4277. They are very similar in terms of morphology, size, and luminosity. But NGC\,4264 hosts a fast bar, which nearly extends out to its corotation ($\mathcal{R}=0.9\pm0.2$, \citealt{Cuomo2019}), while the bar embedded in NGC\,4277 is slow and falls short of the corotation ($\mathcal{R}=1.8^{+0.5}_{-0.3}$, \citealt{Buttitta2022}). We focused on these galaxies because their bar pattern speeds are amongst the best-constrained ones obtained with direct measurements through the Tremaine-Weinberg method \citep{TW1984}. We considered both mass-follows-light models and mass models with a spherical halo of DM, which is not tied to the stars, by matching the stellar kinematics obtained with the MUSE integral-field spectrograph and using SDSS images to recover the stellar mass distribution. 

For both galaxies, the best-fitting mass model has a quasi-isothermal halo for which we derived the fraction of dark matter $f_{\rm DM, bar}$ within the bar region. This is the first time that $\mathcal{R}$ is measured along with $f_{\rm DM, bar}$ obtained from dynamical modelling. We found that the inner regions of NGC\,4277 host a larger amount of dark matter ($f_{\rm DM, bar}\sim0.5$) with respect to NGC\,4264 ($f_{\rm DM, bar}\sim0.3$) in agreement with the predictions of theoretical works and the findings of numerical simulations. Indeed, fast bars are expected to live in baryon-dominated discs, whereas slow bars have experienced a strong drag from the dynamical friction due to a dense halo of dark matter. First, \cite{Weinberg1985} predicted that a DM halo with a significant central mass density exerts a dynamical torque on the bar causing its slowdown. Similar results were later confirmed by several works based on N-body simulations \citep{Debattista1998, Debattista2000, Athanassoula2002, Martinez-Valpuesta2017, Petersen2019}. The bar in a less massive DM halo could require a longer timescale to be slowed down. Tailoring numerical simulations to well-studied galaxies like NGC\,4264 and NGC\,4277 will make it possible to track the decrease of the bar pattern speed as a function of the DM content and to predict the present value of $\mathcal{R}$ to be compared with observations.

According to the results of numerical simulations, tidally induced bars are typically slower than those spontaneously formed by internal instabilities \citep{Miwa1998, Martinez-Valpuesta2017, Lokas2018}. \cite{Martinez-Valpuesta2017} found that bars formed after coplanar flybys with massive companions typically have a rotation rate $\mathcal{R}>1.8$. Similarly, large rotation rates characterise the bars formed in the numerical experiments of \citet{Lokas2018}, which explored the case of retrograde encounters of two galaxies with comparable mass. \citet{Buttitta2022} argued that the formation of the slow bar in NGC\,4277 could be induced by the close companion NGC\,4273. NGC\,4277 does not present strong evidence of tidal interaction with the nearby galaxy, therefore it is not possible to confirm this scenario. However, we can conclude that the large DM fraction in its inner regions influenced the evolution of the bar driving it to an even slower regime. On the other hand, \citet{Cuomo2019} pointed out that the mild interaction with NGC\,4261 favours the spontaneous formation of the fast bar of NGC\,4264. We conclude that the low DM fraction in the bar-dominated region was not enough to efficiently slow down the bar to the slow regime. 

Following theoretical results \citep[e.g.][]{Athanassoula2013, Sellwood2014} angular momentum is exchanged at resonances, but it is emitted at the inner resonances and absorbed at the outer ones. However, the estimate of the amount of angular momentum exchanged at the resonances would require tracking the temporal evolution of the bar parameters of NGC\,4264 and NGC\,4277 through tailored N-body simulations. Nevertheless, it is possible to estimate the amount of DM enclosed within the corotation radius. The location of the corotation and bar radius of NGC\,4264 are consistent with each other within errors, whereas the corotation radius of NGC\,4277 is about twice as large as the bar radius. For NGC\,4264, the DM fraction enclosed within the corotation radius ($f_{\rm DM, cor} = 0.29\pm0.04$) is the same within errors as that enclosed within the bar radius ($f_{\rm DM, bar} = 0.33\pm0.04$). On the contrary, for NGC\,4277 the DM fraction enclosed within the corotation radius ($f_{\rm DM, cor} = 0.72\pm0.03$) is much larger than that inside the bar radius ($f_{\rm DM, bar} = 0.53\pm0.02$) further supporting our conclusions.

A systematic application of the JAM algorithm to simulated barred galaxies with bars of different orientations, lengths, strengths, and pattern speeds is required to extend the parameter space explored by \citet{Lablanche2012}. This is needed to understand the uncertainties and biases on the dynamical parameters, including the DM content in the bar-dominated region, for a given galaxy configuration. In particular, the regimes of weakly-barred galaxies and dwarf-barred galaxies, which give a major contribution to the galaxy population, are yet to be explored. 

This is a crucial step to extend this dynamical analysis to all the barred galaxies with a measured bar rotation rate from integral-field spectroscopic data, like those targeted by CALIFA \citep{CALIFA, Aguerri2015, Cuomo2019b}, MANGA \citep{MANGA, Guo2019, Garma-Oehmichen2020, Garma-Oehmichen2022}, and PHANGS-MUSE \citep{PHANGSMUSE, Williams2021}. With a large set of modelled galaxies, it will be possible to look for a quantitative relationship between $\mathcal{R}$ and $f_{\rm DM, bar}$ with the aim of using $\mathcal{R}$ as a diagnostic of the DM content in galaxy disks. 

Finally, NGC\,4264 and NGC\,4277 are amongst the barred galaxies with the best-studied bar properties in terms of surface-brightness distribution, stellar kinematics, and mass modelling. For this reason, they are ideal candidates to be adopted as a test bench in applying the orbit-superposition Schwarzschild dynamical models, which have been recently developed for tumbling bars by \citet{Vasiliev2020} and \citet{Tahmasebzadeh2022} but which have been applied to date only to mock galaxies created from N-body simulations.

\section*{Acknowledgements}
We thank the referee for a very constructive report that helped to improve the manuscript. CB, EMC, and AP are supported by MIUR grant PRIN 2017 20173ML3WW-001 and Padua University grants DOR2019-2021. JALA and LC are supported by the Spanish Ministerio de Ciencia e Innovaci\'on y Universidades by the grants PID2020-119342GB-I00 and PGC2018-093499-B-I00, respectively. LC acknowledges financial support from Comunidad de Madrid under Atracci\'on de Talento grant 2018-T2/TIC-11612. VC is supported by Fondecyt Postdoctoral programme 3220206-2022 and by ESO-Chile Joint Committee programme ORP060/19. The analysis in this paper was realised with the use of the Python packages {\sc astropy} \citep{Astropy}, {\sc matplotlib} \citep{Matplotlib}, {\sc numpy} \citep{NumPy}, and {\sc scipy} \citep{SciPy}. The MGE decomposition and JAM dynamical models were performed with the Python packages {\sc mgefit} \citep{MGEPYTHON} and {\sc jampy} \citep{Cappellari2008}.

\section*{Data Availability}
The data underlying this article will be shared on reasonable request to CB.

%%%%%%%%%%%%%%%%%%%% REFERENCES %%%%%%%%%%%%%%%%%%

% The best way to enter references is to use BibTeX:

\bibliographystyle{mnras}
\bibliography{biblio.bib} 

\begin{thebibliography}{}
\makeatletter
\relax
\def\mn@urlcharsother{\let\do\@makeother \do\$\do\&\do\#\do\^\do\_\do\%\do\~}
\def\mn@doi{\begingroup\mn@urlcharsother \@ifnextchar [ {\mn@doi@}
  {\mn@doi@[]}}
\def\mn@doi@[#1]#2{\def\@tempa{#1}\ifx\@tempa\@empty \href
  {http://dx.doi.org/#2} {doi:#2}\else \href {http://dx.doi.org/#2} {#1}\fi
  \endgroup}
\def\mn@eprint#1#2{\mn@eprint@#1:#2::\@nil}
\def\mn@eprint@arXiv#1{\href {http://arxiv.org/abs/#1} {{\tt arXiv:#1}}}
\def\mn@eprint@dblp#1{\href {http://dblp.uni-trier.de/rec/bibtex/#1.xml}
  {dblp:#1}}
\def\mn@eprint@#1:#2:#3:#4\@nil{\def\@tempa {#1}\def\@tempb {#2}\def\@tempc
  {#3}\ifx \@tempc \@empty \let \@tempc \@tempb \let \@tempb \@tempa \fi \ifx
  \@tempb \@empty \def\@tempb {arXiv}\fi \@ifundefined
  {mn@eprint@\@tempb}{\@tempb:\@tempc}{\expandafter \expandafter \csname
  mn@eprint@\@tempb\endcsname \expandafter{\@tempc}}}

\bibitem[\protect\citeauthoryear{{Abolfathi} et~al.,}{{Abolfathi}
  et~al.}{2018}]{SDSSDR14}
{Abolfathi} B.,  et~al., 2018, \mn@doi [ApJS] {10.3847/1538-4365/aa9e8a}, \href
  {https://ui.adsabs.harvard.edu/abs/2018ApJS..235...42A} {235, 42}

\bibitem[\protect\citeauthoryear{{Aguerri}, {Mu{\~n}oz-Tu{\~n}{\'o}n}  \&
  {Varela}}{{Aguerri} et~al.}{2000}]{Aguerri2000}
{Aguerri} J.~A.~L.,  {Mu{\~n}oz-Tu{\~n}{\'o}n} C.,   {Varela} A.~M. e.~a.,
  2000, A\&A, \href {http://adsabs.harvard.edu/abs/2000A\%26A...361..841A}
  {361, 841}

\bibitem[\protect\citeauthoryear{{Aguerri}, {Debattista}  \&
  {Corsini}}{{Aguerri} et~al.}{2003}]{Aguerri2003}
{Aguerri} J.~A.~L.,  {Debattista} V.~P.,   {Corsini} E.~M.,  2003, MNRAS, \href
  {https://ui.adsabs.harvard.edu/abs/2003MNRAS.338..465A} {338, 465}

\bibitem[\protect\citeauthoryear{{Aguerri}, {M{\'e}ndez-Abreu}  \&
  {Corsini}}{{Aguerri} et~al.}{2009}]{Aguerri2009}
{Aguerri} J.~A.~L.,  {M{\'e}ndez-Abreu} J.,   {Corsini} E.~M.,  2009, A\&A,
  \href {http://adsabs.harvard.edu/abs/2009A\%26A...495..491A} {495, 491}

\bibitem[\protect\citeauthoryear{{Aguerri} et~al.,}{{Aguerri}
  et~al.}{2015}]{Aguerri2015}
{Aguerri} J.~A.~L.,  et~al., 2015, A\&A, \href
  {https://ui.adsabs.harvard.edu/abs/2015A&A...576A.102A} {576, A102}

\bibitem[\protect\citeauthoryear{{Algorry} et~al.,}{{Algorry}
  et~al.}{2017}]{Algorry2017}
{Algorry} D.~G.,  et~al., 2017, \mn@doi [\mnras] {10.1093/mnras/stx1008}, \href
  {https://ui.adsabs.harvard.edu/abs/2017MNRAS.469.1054A} {469, 1054}

\bibitem[\protect\citeauthoryear{{Astropy Collaboration} et~al.,}{{Astropy
  Collaboration} et~al.}{2018}]{Astropy}
{Astropy Collaboration} et~al., 2018, \mn@doi [\aj] {10.3847/1538-3881/aabc4f},
  \href {https://ui.adsabs.harvard.edu/abs/2018AJ....156..123A} {156, 123}

\bibitem[\protect\citeauthoryear{{Athanassoula}}{{Athanassoula}}{1992}]{Athanassoula1992}
{Athanassoula} E.,  1992, \mn@doi [MNRAS] {10.1093/mnras/259.2.345}, \href
  {http://adsabs.harvard.edu/abs/1992MNRAS.259..345A} {259, 345}

\bibitem[\protect\citeauthoryear{{Athanassoula} \& {Misiriotis}}{{Athanassoula}
  \& {Misiriotis}}{2002}]{Athanassoula2002}
{Athanassoula} E.,  {Misiriotis} A.,  2002, MNRAS, \href
  {http://adsabs.harvard.edu/abs/2002MNRAS.330...35A} {330, 35}

\bibitem[\protect\citeauthoryear{{Athanassoula}, {Machado}  \&
  {Rodionov}}{{Athanassoula} et~al.}{2013}]{Athanassoula2013}
{Athanassoula} E.,  {Machado} R. E.~G.,   {Rodionov} S.~A.,  2013, MNRAS, \href
  {https://ui.adsabs.harvard.edu/abs/2013MNRAS.429.1949A} {429, 1949}

\bibitem[\protect\citeauthoryear{{Barnab{\`e}} et~al.,}{{Barnab{\`e}}
  et~al.}{2012}]{Barnabe2012}
{Barnab{\`e}} M.,  et~al., 2012, \mn@doi [\mnras]
  {10.1111/j.1365-2966.2012.20934.x}, \href
  {https://ui.adsabs.harvard.edu/abs/2012MNRAS.423.1073B} {423, 1073}

\bibitem[\protect\citeauthoryear{{Binney} \& {Tremaine}}{{Binney} \&
  {Tremaine}}{1987}]{Binney1987}
{Binney} J.,  {Tremaine} S.,  1987, {Galactic Dynamics}.
Princeton: Princeton University Press

\bibitem[\protect\citeauthoryear{{Bundy} et~al.,}{{Bundy} et~al.}{2015}]{MANGA}
{Bundy} K.,  et~al., 2015, ApJ, \href
  {https://ui.adsabs.harvard.edu/abs/2015ApJ...798....7B} {798, 7}

\bibitem[\protect\citeauthoryear{{Buta} et~al.,}{{Buta}
  et~al.}{2015}]{Buta2015}
{Buta} R.~J.,  et~al., 2015, ApJS, \href
  {https://ui.adsabs.harvard.edu/abs/2015ApJS..217...32B} {217, 32}

\bibitem[\protect\citeauthoryear{{Buttitta} et~al.,}{{Buttitta}
  et~al.}{2022}]{Buttitta2022}
{Buttitta} C.,  et~al., 2022, \mn@doi [\aap] {10.1051/0004-6361/202244297},
  \href {https://ui.adsabs.harvard.edu/abs/2022A&A...664L..10B} {664, L10}

\bibitem[\protect\citeauthoryear{{Cappellari}}{{Cappellari}}{2002}]{MGEPYTHON}
{Cappellari} M.,  2002, \mn@doi [\mnras] {10.1046/j.1365-8711.2002.05412.x},
  \href {https://ui.adsabs.harvard.edu/abs/2002MNRAS.333..400C} {333, 400}

\bibitem[\protect\citeauthoryear{{Cappellari}}{{Cappellari}}{2008}]{Cappellari2008}
{Cappellari} M.,  2008, \mn@doi [MNRAS] {10.1111/j.1365-2966.2008.13754.x},
  \href {https://ui.adsabs.harvard.edu/abs/2008MNRAS.390...71C} {390, 71}

\bibitem[\protect\citeauthoryear{{Cappellari}}{{Cappellari}}{2020}]{Cappellari2020}
{Cappellari} M.,  2020, \mn@doi [MNRAS] {10.1093/mnras/staa959}, \href
  {https://ui.adsabs.harvard.edu/abs/2020MNRAS.494.4819C} {494, 4819}

\bibitem[\protect\citeauthoryear{{Cappellari} \& {Copin}}{{Cappellari} \&
  {Copin}}{2003}]{Cappellari2003}
{Cappellari} M.,  {Copin} Y.,  2003, MNRAS, \href
  {https://ui.adsabs.harvard.edu/abs/2003MNRAS.342..345C} {342, 345}

\bibitem[\protect\citeauthoryear{{Cappellari} \& {Emsellem}}{{Cappellari} \&
  {Emsellem}}{2004}]{Cappellari2004}
{Cappellari} M.,  {Emsellem} E.,  2004, PASP, \href
  {https://ui.adsabs.harvard.edu/abs/2004PASP..116..138C} {116, 138}

\bibitem[\protect\citeauthoryear{{Cappellari} et~al.,}{{Cappellari}
  et~al.}{2011}]{Cappellari2011}
{Cappellari} M.,  et~al., 2011, \mn@doi [\mnras]
  {10.1111/j.1365-2966.2010.18174.x}, \href
  {https://ui.adsabs.harvard.edu/abs/2011MNRAS.413..813C} {413, 813}

\bibitem[\protect\citeauthoryear{{Cappellari} et~al.,}{{Cappellari}
  et~al.}{2013}]{Cappellari2013}
{Cappellari} M.,  et~al., 2013, \mn@doi [\mnras] {10.1093/mnras/stt562}, \href
  {https://ui.adsabs.harvard.edu/abs/2013MNRAS.432.1709C} {432, 1709}

\bibitem[\protect\citeauthoryear{{Chilingarian} \& {Zolotukhin}}{{Chilingarian}
  \& {Zolotukhin}}{2012}]{Chilingarian2012}
{Chilingarian} I.~V.,  {Zolotukhin} I.~Y.,  2012, \mn@doi [MNRAS]
  {10.1111/j.1365-2966.2011.19837.x}, \href
  {https://ui.adsabs.harvard.edu/abs/2012MNRAS.419.1727C} {419, 1727}

\bibitem[\protect\citeauthoryear{{Corsini}}{{Corsini}}{2011}]{Corsini2011}
{Corsini} E.~M.,  2011, Mem. Soc. Astron. It. Sup., \href
  {http://adsabs.harvard.edu/abs/2011MSAIS..18...23C} {18, 23}

\bibitem[\protect\citeauthoryear{{Cuomo} et~al.,}{{Cuomo}
  et~al.}{2019a}]{Cuomo2019}
{Cuomo} V.,  et~al., 2019a, \mn@doi [MNRAS] {10.1093/mnras/stz1943}, \href
  {https://ui.adsabs.harvard.edu/abs/2019MNRAS.488.4972C} {488, 4972}

\bibitem[\protect\citeauthoryear{{Cuomo}, {Lopez Aguerri}, {Corsini},
  {Debattista}, {M{\'e}ndez-Abreu}  \& {Pizzella}}{{Cuomo}
  et~al.}{2019b}]{Cuomo2019b}
{Cuomo} V.,  {Lopez Aguerri} J.~A.,  {Corsini} E.~M.,  {Debattista} V.~P.,
  {M{\'e}ndez-Abreu} J.,   {Pizzella} A.,  2019b, \mn@doi [A\&A]
  {10.1051/0004-6361/201936415}, \href
  {https://ui.adsabs.harvard.edu/abs/2019A&A...632A..51C} {632, A51}

\bibitem[\protect\citeauthoryear{{Cuomo}, {Aguerri}, {Corsini}  \&
  {Debattista}}{{Cuomo} et~al.}{2020}]{Cuomo2020}
{Cuomo} V.,  {Aguerri} J.~A.~L.,  {Corsini} E.~M.,   {Debattista} V.~P.,  2020,
  \mn@doi [A\&A] {10.1051/0004-6361/202037945}, \href
  {https://ui.adsabs.harvard.edu/abs/2020A&A...641A.111C} {641, A111}

\bibitem[\protect\citeauthoryear{{Debattista} \& {Sellwood}}{{Debattista} \&
  {Sellwood}}{1998}]{Debattista1998}
{Debattista} V.~P.,  {Sellwood} J.~A.,  1998, ApJL, \href
  {https://ui.adsabs.harvard.edu/abs/1998ApJ...493L...5D} {493, L5}

\bibitem[\protect\citeauthoryear{{Debattista} \& {Sellwood}}{{Debattista} \&
  {Sellwood}}{2000}]{Debattista2000}
{Debattista} V.~P.,  {Sellwood} J.~A.,  2000, ApJ, \href
  {http://adsabs.harvard.edu/abs/2000ApJ...543..704D} {543, 704}

\bibitem[\protect\citeauthoryear{{Debattista}, {Mayer}, {Carollo}, {Moore},
  {Wadsley}  \& {Quinn}}{{Debattista} et~al.}{2006}]{Debattista2006}
{Debattista} V.~P.,  {Mayer} L.,  {Carollo} C.~M.,  {Moore} B.,  {Wadsley} J.,
   {Quinn} T.,  2006, ApJ, \href
  {https://ui.adsabs.harvard.edu/abs/2006ApJ...645..209D} {645, 209}

\bibitem[\protect\citeauthoryear{{Emsellem} et~al.,}{{Emsellem}
  et~al.}{2022}]{PHANGSMUSE}
{Emsellem} E.,  et~al., 2022, \aap, \href
  {https://ui.adsabs.harvard.edu/abs/2022A&A...659A.191E} {659, A191}

\bibitem[\protect\citeauthoryear{{Fixsen}, {Cheng}, {Gales}, {Mather}, {Shafer}
   \& {Wright}}{{Fixsen} et~al.}{1996}]{Fixsen1996}
{Fixsen} D.~J.,  {Cheng} E.~S.,  {Gales} J.~M.,  {Mather} J.~C.,  {Shafer}
  R.~A.,   {Wright} E.~L.,  1996, ApJ, \href
  {https://ui.adsabs.harvard.edu/abs/1996ApJ...473..576F} {473, 576}

\bibitem[\protect\citeauthoryear{{Fragkoudi}, {Athanassoula}  \&
  {Bosma}}{{Fragkoudi} et~al.}{2016}]{Fragkoudi2016}
{Fragkoudi} F.,  {Athanassoula} E.,   {Bosma} A.,  2016, \mn@doi [MNRAS]
  {10.1093/mnrasl/slw120}, \href
  {https://ui.adsabs.harvard.edu/abs/2016MNRAS.462L..41F} {462, L41}

\bibitem[\protect\citeauthoryear{{Fragkoudi}, {Grand}, {Pakmor}, {Springel},
  {White}, {Marinacci}, {Gomez}  \& {Navarro}}{{Fragkoudi}
  et~al.}{2021}]{Fragkoudi2021}
{Fragkoudi} F.,  {Grand} R.~J.~J.,  {Pakmor} R.,  {Springel} V.,  {White}
  S.~D.~M.,  {Marinacci} F.,  {Gomez} F.~A.,   {Navarro} J.~F.,  2021, \mn@doi
  [\aap] {10.1051/0004-6361/202140320}, \href
  {https://ui.adsabs.harvard.edu/abs/2021A&A...650L..16F} {650, L16}

\bibitem[\protect\citeauthoryear{{Frosst}, {Courteau}, {Arora}, {Stone},
  {Macci{\`o}}  \& {Blank}}{{Frosst} et~al.}{2022}]{Frosst2022}
{Frosst} M.,  {Courteau} S.,  {Arora} N.,  {Stone} C.,  {Macci{\`o}} A.~V.,
  {Blank} M.,  2022, \mn@doi [MNRAS] {10.1093/mnras/stac1497}, \href
  {https://ui.adsabs.harvard.edu/abs/2022MNRAS.514.3510F} {514, 3510}

\bibitem[\protect\citeauthoryear{{Garma-Oehmichen}, {Cano-D{\'\i}az},
  {Hern{\'a}ndez-Toledo}, {Aquino-Ort{\'\i}z}, {Valenzuela}, {Aguerri},
  {S{\'a}nchez}  \& {Merrifield}}{{Garma-Oehmichen}
  et~al.}{2020}]{Garma-Oehmichen2020}
{Garma-Oehmichen} L.,  {Cano-D{\'\i}az} M.,  {Hern{\'a}ndez-Toledo} H.,
  {Aquino-Ort{\'\i}z} E.,  {Valenzuela} O.,  {Aguerri} J.~A.~L.,  {S{\'a}nchez}
  S.~F.,   {Merrifield} M.,  2020, \mn@doi [MNRAS] {10.1093/mnras/stz3101},
  \href {https://ui.adsabs.harvard.edu/abs/2020MNRAS.491.3655G} {491, 3655}

\bibitem[\protect\citeauthoryear{{Garma-Oehmichen} et~al.,}{{Garma-Oehmichen}
  et~al.}{2022}]{Garma-Oehmichen2022}
{Garma-Oehmichen} L.,  et~al., 2022, \mn@doi [\mnras] {10.1093/mnras/stac3069},
  \href {https://ui.adsabs.harvard.edu/abs/2022MNRAS.tmp.2873G} {}

\bibitem[\protect\citeauthoryear{{Gerhard}}{{Gerhard}}{1993}]{Gerhard1993}
{Gerhard} O.~E.,  1993, MNRAS, \href
  {https://ui.adsabs.harvard.edu/abs/1993MNRAS.265..213G} {265, 213}

\bibitem[\protect\citeauthoryear{{Gerhard} \& {Binney}}{{Gerhard} \&
  {Binney}}{1996}]{Gerhard1996}
{Gerhard} O.~E.,  {Binney} J.~J.,  1996, \mn@doi [\mnras]
  {10.1093/mnras/279.3.993}, \href
  {https://ui.adsabs.harvard.edu/abs/1996MNRAS.279..993G} {279, 993}

\bibitem[\protect\citeauthoryear{{Gottloeber}, {Hoffman}  \&
  {Yepes}}{{Gottloeber} et~al.}{2010}]{CLUES2010}
{Gottloeber} S.,  {Hoffman} Y.,   {Yepes} G.,  2010, in {Wagner} S.,
  {Steinmetz} M.,  {Bode} A.,   {Muller} M.,  eds, {High Performance Computing
  in Science and Engineering}. Springer.
p.~309

\bibitem[\protect\citeauthoryear{{Grand} et~al.,}{{Grand}
  et~al.}{2017}]{AURIGA2017}
{Grand} R. J.~J.,  et~al., 2017, \mn@doi [\mnras] {10.1093/mnras/stx071}, \href
  {https://ui.adsabs.harvard.edu/abs/2017MNRAS.467..179G} {467, 179}

\bibitem[\protect\citeauthoryear{{Guo}, {Mao}, {Athanassoula}, {Li}, {Ge},
  {Long}, {Merrifield}  \& {Masters}}{{Guo} et~al.}{2019}]{Guo2019}
{Guo} R.,  {Mao} S.,  {Athanassoula} E.,  {Li} H.,  {Ge} J.,  {Long} R.~J.,
  {Merrifield} M.,   {Masters} K.,  2019, MNRAS, \href
  {https://ui.adsabs.harvard.edu/abs/2019MNRAS.482.1733G} {482, 1733}

\bibitem[\protect\citeauthoryear{Hunter}{Hunter}{2007}]{Matplotlib}
Hunter J.~D.,  2007, \mn@doi [Computing in Science \& Engineering]
  {10.1109/MCSE.2007.55}, 9, 90

\bibitem[\protect\citeauthoryear{{Kalinova} et~al.,}{{Kalinova}
  et~al.}{2017}]{Kalinova2017}
{Kalinova} V.,  et~al., 2017, \mn@doi [MNRAS] {10.1093/mnras/stx901}, \href
  {https://ui.adsabs.harvard.edu/abs/2017MNRAS.469.2539K} {469, 2539}

\bibitem[\protect\citeauthoryear{{Kim} et~al.,}{{Kim} et~al.}{2014}]{Kim2014}
{Kim} S.,  et~al., 2014, ApJS, \href
  {https://ui.adsabs.harvard.edu/abs/2014ApJS..215...22K} {215, 22}

\bibitem[\protect\citeauthoryear{{Klypin}, {Trujillo-Gomez}  \&
  {Primack}}{{Klypin} et~al.}{2011}]{Klypin2011}
{Klypin} A.~A.,  {Trujillo-Gomez} S.,   {Primack} J.,  2011, \mn@doi [\apj]
  {10.1088/0004-637X/740/2/102}, \href
  {https://ui.adsabs.harvard.edu/abs/2011ApJ...740..102K} {740, 102}

\bibitem[\protect\citeauthoryear{{Kormendy} \& {Ho}}{{Kormendy} \&
  {Ho}}{2013}]{Kormendy2013}
{Kormendy} J.,  {Ho} L.~C.,  2013, \mn@doi [\araa]
  {10.1146/annurev-astro-082708-101811}, \href
  {https://ui.adsabs.harvard.edu/abs/2013ARA&A..51..511K} {51, 511}

\bibitem[\protect\citeauthoryear{{Krajnovi{\'c}}, {Cappellari}, {Emsellem},
  {McDermid}  \& {de Zeeuw}}{{Krajnovi{\'c}} et~al.}{2005}]{Krajnovic2005}
{Krajnovi{\'c}} D.,  {Cappellari} M.,  {Emsellem} E.,  {McDermid} R.~M.,   {de
  Zeeuw} P.~T.,  2005, \mn@doi [\mnras] {10.1111/j.1365-2966.2005.08715.x},
  \href {https://ui.adsabs.harvard.edu/abs/2005MNRAS.357.1113K} {357, 1113}

\bibitem[\protect\citeauthoryear{{Lablanche} et~al.,}{{Lablanche}
  et~al.}{2012}]{Lablanche2012}
{Lablanche} P.-Y.,  et~al., 2012, \mn@doi [MNRAS]
  {10.1111/j.1365-2966.2012.21343.x}, \href
  {https://ui.adsabs.harvard.edu/abs/2012MNRAS.424.1495L} {424, 1495}

\bibitem[\protect\citeauthoryear{{Laurikainen}, {Salo}, {Buta}  \&
  {Knapen}}{{Laurikainen} et~al.}{2007}]{Laurikainen2007}
{Laurikainen} E.,  {Salo} H.,  {Buta} R.,   {Knapen} J.~H.,  2007, \mn@doi
  [MNRAS] {10.1111/j.1365-2966.2007.12299.x}, \href
  {https://ui.adsabs.harvard.edu/abs/2007MNRAS.381..401L} {381, 401}

\bibitem[\protect\citeauthoryear{{{\L}okas}}{{{\L}okas}}{2018}]{Lokas2018}
{{\L}okas} E.~L.,  2018, ApJ, 857, 6

\bibitem[\protect\citeauthoryear{{Marioni}, {Abadi}, {Gottl{\"o}ber}  \&
  {Yepes}}{{Marioni} et~al.}{2022}]{Marioni2022}
{Marioni} O.~F.,  {Abadi} M.~G.,  {Gottl{\"o}ber} S.,   {Yepes} G.,  2022,
  \mnras, 511, 2423

\bibitem[\protect\citeauthoryear{{Martinez-Valpuesta}, {Aguerri},
  {Gonz{\'a}lez-Garc{\'{\i}}a}, {Dalla Vecchia}  \&
  {Stringer}}{{Martinez-Valpuesta} et~al.}{2017}]{Martinez-Valpuesta2017}
{Martinez-Valpuesta} I.,  {Aguerri} J.~A.~L.,  {Gonz{\'a}lez-Garc{\'{\i}}a}
  A.~C.,  {Dalla Vecchia} C.,   {Stringer} M.,  2017, MNRAS, \href
  {http://adsabs.harvard.edu/abs/2017MNRAS.464.1502M} {464, 1502}

\bibitem[\protect\citeauthoryear{{M}c{K}inney}{{M}c{K}inney}{2010}]{NumPy}
{M}c{K}inney 2010, in {S}t\'efan van~der {W}alt {J}arrod {M}illman eds,
  {P}roceedings of the 9th {P}ython in {S}cience {C}onference. pp 56 -- 61,
  \mn@doi{10.25080/Majora-92bf1922-00a}

\bibitem[\protect\citeauthoryear{{M{\'e}ndez-Abreu} et~al.,}{{M{\'e}ndez-Abreu}
  et~al.}{2017}]{MendezAbreu2017}
{M{\'e}ndez-Abreu} J.,  et~al., 2017, \aap, \href
  {http://adsabs.harvard.edu/abs/2017A\%26A...598A..32M} {598, A32}

\bibitem[\protect\citeauthoryear{{Miwa} \& {Noguchi}}{{Miwa} \&
  {Noguchi}}{1998}]{Miwa1998}
{Miwa} T.,  {Noguchi} M.,  1998, \mn@doi [\apj] {10.1086/305611}, \href
  {https://ui.adsabs.harvard.edu/abs/1998ApJ...499..149M} {499, 149}

\bibitem[\protect\citeauthoryear{{Navarro}, {Frenk}  \& {White}}{{Navarro}
  et~al.}{1995}]{NFWDM1995}
{Navarro} J.~F.,  {Frenk} C.~S.,   {White} S. D.~M.,  1995, \mn@doi [MNRAS]
  {10.1093/mnras/275.1.56}, \href
  {https://ui.adsabs.harvard.edu/abs/1995MNRAS.275...56N} {275, 56}

\bibitem[\protect\citeauthoryear{{Nelson} et~al.,}{{Nelson}
  et~al.}{2018}]{Nelson2018}
{Nelson} D.,  et~al., 2018, \mn@doi [\mnras] {10.1093/mnras/stx3040}, \href
  {https://ui.adsabs.harvard.edu/abs/2018MNRAS.475..624N} {475, 624}

\bibitem[\protect\citeauthoryear{{Noguchi}}{{Noguchi}}{1987}]{Noguchi1987}
{Noguchi} M.,  1987, MNRAS, \href
  {https://ui.adsabs.harvard.edu/abs/1987MNRAS.228..635N} {228, 635}

\bibitem[\protect\citeauthoryear{{Noordermeer}, {van der Hulst}, {Sancisi},
  {Swaters}  \& {van Albada}}{{Noordermeer} et~al.}{2007}]{Noordermeer2007}
{Noordermeer} E.,  {van der Hulst} J.~M.,  {Sancisi} R.,  {Swaters} R.~S.,
  {van Albada} T.~S.,  2007, \mn@doi [MNRAS]
  {10.1111/j.1365-2966.2007.11533.x}, \href
  {https://ui.adsabs.harvard.edu/abs/2007MNRAS.376.1513N} {376, 1513}

\bibitem[\protect\citeauthoryear{{Pagotto}, {Corsini}, {Sarzi}, {Pagani},
  {Dalla Bont{\`a}}, {Morelli}  \& {Pizzella}}{{Pagotto}
  et~al.}{2019}]{Pagotto2019}
{Pagotto} I.,  {Corsini} E.~M.,  {Sarzi} M.,  {Pagani} B.,  {Dalla Bont{\`a}}
  E.,  {Morelli} L.,   {Pizzella} A.,  2019, \mn@doi [MNRAS]
  {10.1093/mnras/sty2918}, \href
  {https://ui.adsabs.harvard.edu/abs/2019MNRAS.483...57P} {483, 57}

\bibitem[\protect\citeauthoryear{{Petersen}, {Weinberg}  \& {Katz}}{{Petersen}
  et~al.}{2019}]{Petersen2019}
{Petersen} M.~S.,  {Weinberg} M.~D.,   {Katz} N.,  2019, \mn@doi [\mnras]
  {10.1093/mnras/stz2824}, \href
  {https://ui.adsabs.harvard.edu/abs/2019MNRAS.490.3616P} {490, 3616}

\bibitem[\protect\citeauthoryear{{Pillepich} et~al.,}{{Pillepich}
  et~al.}{2018}]{Pillepich2018}
{Pillepich} A.,  et~al., 2018, \mn@doi [\mnras] {10.1093/mnras/stx3112}, \href
  {https://ui.adsabs.harvard.edu/abs/2018MNRAS.475..648P} {475, 648}

\bibitem[\protect\citeauthoryear{Portail, Gerhard, Wegg  \& Ness}{Portail
  et~al.}{2016}]{Portail2017}
Portail M.,  Gerhard O.,  Wegg C.,   Ness M.,  2016, \mn@doi [Monthly Notices
  of the Royal Astronomical Society] {10.1093/mnras/stw2819}, 465, 1621

\bibitem[\protect\citeauthoryear{{Prugniel} \& {Soubiran}}{{Prugniel} \&
  {Soubiran}}{2001}]{Prugniel2001}
{Prugniel} P.,  {Soubiran} C.,  2001, A\&A, \href
  {https://ui.adsabs.harvard.edu/abs/2001A&A...369.1048P} {369, 1048}

\bibitem[\protect\citeauthoryear{{Roshan}, {Ghafourian}, {Kashfi}, {Banik},
  {Haslbauer}, {Cuomo}, {Famaey}  \& {Kroupa}}{{Roshan}
  et~al.}{2021}]{Roshan2021}
{Roshan} M.,  {Ghafourian} N.,  {Kashfi} T.,  {Banik} I.,  {Haslbauer} M.,
  {Cuomo} V.,  {Famaey} B.,   {Kroupa} P.,  2021, \mn@doi [MNRAS]
  {10.1093/mnras/stab2553}, \href
  {https://ui.adsabs.harvard.edu/abs/2021MNRAS.508..926R} {508, 926}

\bibitem[\protect\citeauthoryear{{Rybicki}}{{Rybicki}}{1987}]{Rybicki1987}
{Rybicki} G.~B.,  1987, in {de Zeeuw} P.~T.,  ed., ~ Vol. 127, Structure and
  Dynamics of Elliptical Galaxies. p.~397

\bibitem[\protect\citeauthoryear{{S{\'a}nchez} et~al.,}{{S{\'a}nchez}
  et~al.}{2012}]{CALIFA}
{S{\'a}nchez} S.~F.,  et~al., 2012, A\&A, \href
  {https://ui.adsabs.harvard.edu/abs/2012A&A...538A...8S} {538, A8}

\bibitem[\protect\citeauthoryear{{Sarzi} et~al.,}{{Sarzi}
  et~al.}{2006}]{Sarzi2006}
{Sarzi} M.,  et~al., 2006, MNRAS, \href
  {https://ui.adsabs.harvard.edu/abs/2006MNRAS.366.1151S} {366, 1151}

\bibitem[\protect\citeauthoryear{{Schaye} et~al.,}{{Schaye}
  et~al.}{2015}]{EAGLE2015}
{Schaye} J.,  et~al., 2015, \mn@doi [\mnras] {10.1093/mnras/stu2058}, \href
  {https://ui.adsabs.harvard.edu/abs/2015MNRAS.446..521S} {446, 521}

\bibitem[\protect\citeauthoryear{{Schlafly} \& {Finkbeiner}}{{Schlafly} \&
  {Finkbeiner}}{2011}]{Schlafly2011}
{Schlafly} E.~F.,  {Finkbeiner} D.~P.,  2011, \mn@doi [ApJ]
  {10.1088/0004-637X/737/2/103}, \href
  {https://ui.adsabs.harvard.edu/abs/2011ApJ...737..103S} {737, 103}

\bibitem[\protect\citeauthoryear{{Schmitt}}{{Schmitt}}{2001}]{Schmitt2001}
{Schmitt} H.~R.,  2001, \mn@doi [\aj] {10.1086/323547}, \href
  {https://ui.adsabs.harvard.edu/abs/2001AJ....122.2243S} {122, 2243}

\bibitem[\protect\citeauthoryear{{Scott} et~al.,}{{Scott}
  et~al.}{2013}]{Scott2013}
{Scott} N.,  et~al., 2013, \mn@doi [\mnras] {10.1093/mnras/sts422}, \href
  {https://ui.adsabs.harvard.edu/abs/2013MNRAS.432.1894S} {432, 1894}

\bibitem[\protect\citeauthoryear{{Sellwood}}{{Sellwood}}{1981}]{Sellwood1981}
{Sellwood} J.~A.,  1981, A\&A, \href
  {https://ui.adsabs.harvard.edu/abs/1981A&A....99..362S} {99, 362}

\bibitem[\protect\citeauthoryear{{Sellwood}}{{Sellwood}}{2014}]{Sellwood2014}
{Sellwood} J.~A.,  2014, \mn@doi [Reviews of Modern Physics]
  {10.1103/RevModPhys.86.1}, \href
  {https://ui.adsabs.harvard.edu/abs/2014RvMP...86....1S} {86, 1}

\bibitem[\protect\citeauthoryear{{Tahmasebzadeh}, {Zhu}, {Shen}, {Gerhard}  \&
  {Ven}}{{Tahmasebzadeh} et~al.}{2022}]{Tahmasebzadeh2022}
{Tahmasebzadeh} B.,  {Zhu} L.,  {Shen} J.,  {Gerhard} O.,   {Ven} G. v.~d.,
  2022, \mn@doi [\apj] {10.3847/1538-4357/ac9df6}, \href
  {https://ui.adsabs.harvard.edu/abs/2022ApJ...941..109T} {941, 109}

\bibitem[\protect\citeauthoryear{{Tremaine} \& {Weinberg}}{{Tremaine} \&
  {Weinberg}}{1984}]{TW1984}
{Tremaine} S.,  {Weinberg} M.~D.,  1984, ApJL, \href
  {https://ui.adsabs.harvard.edu/abs/1984ApJ...282L...5T} {282, L5}

\bibitem[\protect\citeauthoryear{{Vasiliev} \& {Valluri}}{{Vasiliev} \&
  {Valluri}}{2020}]{Vasiliev2020}
{Vasiliev} E.,  {Valluri} M.,  2020, \mn@doi [ApJ] {10.3847/1538-4357/ab5fe0},
  \href {https://ui.adsabs.harvard.edu/abs/2020ApJ...889...39V} {889, 39}

\bibitem[\protect\citeauthoryear{{Weinberg}}{{Weinberg}}{1985}]{Weinberg1985}
{Weinberg} M.~D.,  1985, \mn@doi [MNRAS] {10.1093/mnras/213.3.451}, \href
  {https://ui.adsabs.harvard.edu/abs/1985MNRAS.213..451W} {213, 451}

\bibitem[\protect\citeauthoryear{{Williams}, {Bureau}  \&
  {Cappellari}}{{Williams} et~al.}{2009}]{Williams2009}
{Williams} M.~J.,  {Bureau} M.,   {Cappellari} M.,  2009, \mn@doi [MNRAS]
  {10.1111/j.1365-2966.2009.15582.x}, \href
  {https://ui.adsabs.harvard.edu/abs/2009MNRAS.400.1665W} {400, 1665}

\bibitem[\protect\citeauthoryear{{Williams} et~al.,}{{Williams}
  et~al.}{2021}]{Williams2021}
{Williams} T.~G.,  et~al., 2021, \mn@doi [\aj] {10.3847/1538-3881/abe243},
  \href {https://ui.adsabs.harvard.edu/abs/2021AJ....161..185W} {161, 185}

\bibitem[\protect\citeauthoryear{{Willmer}}{{Willmer}}{2018}]{Willmer2018}
{Willmer} C. N.~A.,  2018, \mn@doi [\apjs] {10.3847/1538-4365/aabfdf}, \href
  {https://ui.adsabs.harvard.edu/abs/2018ApJS..236...47W} {236, 47}

\bibitem[\protect\citeauthoryear{{de Lorenzi}, {Debattista}, {Gerhard}  \&
  {Sambhus}}{{de Lorenzi} et~al.}{2007}]{deLorenzi2007}
{de Lorenzi} F.,  {Debattista} V.~P.,  {Gerhard} O.,   {Sambhus} N.,  2007,
  \mn@doi [\mnras] {10.1111/j.1365-2966.2007.11434.x}, \href
  {https://ui.adsabs.harvard.edu/abs/2007MNRAS.376...71D} {376, 71}

\bibitem[\protect\citeauthoryear{{de Vaucouleurs}, {de Vaucouleurs}, {Corwin},
  {Buta}, {Paturel}  \& {Fouqu{\'e}}}{{de Vaucouleurs} et~al.}{1991}]{RC3}
{de Vaucouleurs} G.,  {de Vaucouleurs} A.,  {Corwin} Jr. H.~G.,  {Buta} R.~J.,
  {Paturel} G.,   {Fouqu{\'e}} P.,  1991, {Third Reference Catalogue of Bright
  Galaxies}.
New York: Springer

\bibitem[\protect\citeauthoryear{{de Zeeuw} et~al.,}{{de Zeeuw}
  et~al.}{2002}]{SAURON}
{de Zeeuw} P.~T.,  et~al., 2002, \mn@doi [\mnras]
  {10.1046/j.1365-8711.2002.05059.x}, \href
  {https://ui.adsabs.harvard.edu/abs/2002MNRAS.329..513D} {329, 513}

\bibitem[\protect\citeauthoryear{{van Driel}, {Ragaigne}, {Boselli}, {Donas}
  \& {Gavazzi}}{{van Driel} et~al.}{2000}]{vanDriel2000}
{van Driel} W.,  {Ragaigne} D.,  {Boselli} A.,  {Donas} J.,   {Gavazzi} G.,
  2000, \mn@doi [A\&AS] {10.1051/aas:2000220}, \href
  {https://ui.adsabs.harvard.edu/abs/2000A&AS..144..463V} {144, 463}

\bibitem[\protect\citeauthoryear{{van der Marel} \& {Franx}}{{van der Marel} \&
  {Franx}}{1993}]{Marel1993}
{van der Marel} R.~P.,  {Franx} M.,  1993, ApJ, \href
  {https://ui.adsabs.harvard.edu/abs/1993ApJ...407..525V} {407, 525}

\bibitem[\protect\citeauthoryear{{van der Walt}, {Colbert}  \&
  {Varoquaux}}{{van der Walt} et~al.}{2011}]{SciPy}
{van der Walt} S.,  {Colbert} S.~C.,   {Varoquaux} G.,  2011, Computing in
  Science and Engineering, 13, 22

\makeatother
\end{thebibliography}

% Don't change these lines
\bsp	% typesetting comment
\label{lastpage}

\end{document}